\documentclass[11pt, a4paper]{article}

\usepackage{ifthen}

\newcommand{\MyUniPat}{lsdfgkhjvrkjlhmisdlcjn}

\makeatletter

\newcommand{\NewcommandThreeArgsTwoOpt}[5]
{
\DeclareRobustCommand#1{\@ifnextchar[%
{\csname\expandafter\@gobble\string#1@presq\endcsname}%
{\csname\expandafter\@gobble\string#1@nopresq\endcsname}}
\expandafter\def\csname\expandafter\@gobble\string#1@nopresq\endcsname##1{\@ifnextchar[%
{\csname\expandafter\@gobble\string#1@nopresq@postsq\endcsname[]{##1}}%
{\csname\expandafter\@gobble\string#1@nopresq@nopostsq\endcsname[]{##1}}}
\expandafter\def\csname\expandafter\@gobble\string#1@presq\endcsname[##1]##2{\@ifnextchar[%
{\csname\expandafter\@gobble\string#1@presq@postsq\endcsname[{##1}]{##2}}%
{\csname\expandafter\@gobble\string#1@presq@nopostsq\endcsname[{##1}]{##2}}}
\expandafter\def\csname\expandafter\@gobble\string#1@nopresq@nopostsq\endcsname[##1]##2{#2}
\expandafter\def\csname\expandafter\@gobble\string#1@presq@nopostsq\endcsname[##1]##2{#3}
\expandafter\def\csname\expandafter\@gobble\string#1@nopresq@postsq\endcsname[##1]##2[##3]{#4}
\expandafter\def\csname\expandafter\@gobble\string#1@presq@postsq\endcsname[##1]##2[##3]{#5}
}

\newcommand{\NewcommandTwoArgsLastOpt}[3]
{
\DeclareRobustCommand#1{\csname\expandafter\@gobble\string#1@def\endcsname}
\expandafter\def\csname\expandafter\@gobble\string#1@def\endcsname##1{\@ifnextchar[%
{\csname\expandafter\@gobble\string#1@postsq\endcsname{##1}}%
{\csname\expandafter\@gobble\string#1@nopostsq\endcsname{##1}}}
\expandafter\def\csname\expandafter\@gobble\string#1@nopostsq\endcsname##1{#2}
\expandafter\def\csname\expandafter\@gobble\string#1@postsq\endcsname##1[##2]{#3}
}

\newcommand{\NewcommandFourArgsTwoOpt}[5]
{
\DeclareRobustCommand#1{\@ifnextchar[%
{\csname\expandafter\@gobble\string#1@presq\endcsname}%
{\csname\expandafter\@gobble\string#1@nopresq\endcsname}}
\expandafter\def\csname\expandafter\@gobble\string#1@nopresq\endcsname##1##2{\@ifnextchar[%
{\csname\expandafter\@gobble\string#1@nopresq@postsq\endcsname[]{##1}{##2}}%
{\csname\expandafter\@gobble\string#1@nopresq@nopostsq\endcsname[]{##1}{##2}}}
\expandafter\def\csname\expandafter\@gobble\string#1@presq\endcsname[##1]##2##3{\@ifnextchar[%
{\csname\expandafter\@gobble\string#1@presq@postsq\endcsname[{##1}]{##2}{##3}}%
{\csname\expandafter\@gobble\string#1@presq@nopostsq\endcsname[{##1}]{##2}{##3}}}
\expandafter\def\csname\expandafter\@gobble\string#1@nopresq@nopostsq\endcsname[##1]##2##3{#2}
\expandafter\def\csname\expandafter\@gobble\string#1@presq@nopostsq\endcsname[##1]##2##3{#3}
\expandafter\def\csname\expandafter\@gobble\string#1@nopresq@postsq\endcsname[##1]##2##3[##4]{#4}
\expandafter\def\csname\expandafter\@gobble\string#1@presq@postsq\endcsname[##1]##2##3[##4]{#5}
}

\newcommand{\NewcommandFourArgsTwoOptDefa}[4]
{
\expandafter\newcommand\csname\expandafter\@gobble\string#1@full\endcsname[4]{#4}
\NewcommandFourArgsTwoOpt{#1}
{\csname\expandafter\@gobble\string#1@full\endcsname{#2}{##2}{##3}{#3}}
{\csname\expandafter\@gobble\string#1@full\endcsname{##1}{##2}{##3}{#3}}
{\csname\expandafter\@gobble\string#1@full\endcsname{#2}{##2}{##3}{##4}}
{\csname\expandafter\@gobble\string#1@full\endcsname{##1}{##2}{##3}{##4}}
}

\NewcommandFourArgsTwoOpt{\ifndef}
{\@ifundefined{#2}{#3}{}}%
{\ClassError{My Macros}{The first optional argument is not expected in "\ifndef"}{}}%
{\@ifundefined{#2}{#3}{#4}}%
{\ClassError{My Macros}{The first optional argument is not expected in "\ifndef"}{}}%

\NewcommandThreeArgsTwoOpt{\myusepackage}
{\@ifpackageloaded{#2}{}{\usepackage{#2}}}
{\@ifpackageloaded{#2}{}{\usepackage[#1]{#2}}}
{\@ifpackageloaded{#2}{}{#3 \usepackage{#2}}}
{\@ifpackageloaded{#2}{}{#3 \usepackage[#1]{#2}}}

\makeatother

\myusepackage{amssymb}
\myusepackage{amsmath}%

\myusepackage{amsthm}[ ]%

\myusepackage{tempora}%

\myusepackage[T1,OT2,T2A]{fontenc}

\myusepackage[utf8]{inputenc}
\let\f\relax

\usepackage[shorthands=off,greek,czech,ukrainian,english]{babel}

\let\sq\relax%

\DeclareTextSymbolDefault{\CYRYAT}{OT2}
\DeclareTextSymbolDefault{\cyryat}{OT2}
\DeclareTextSymbolDefault{\CYRFITA}{OT2}
\DeclareTextSymbolDefault{\cyrfita}{OT2}
\DeclareTextSymbolDefault{\CYRIZH}{OT2}
\DeclareTextSymbolDefault{\cyrizh}{OT2}

\myusepackage{soulutf8}%
\setul{1.25pt}{.16pt}%

\myusepackage{latexsym}

\myusepackage{amsfonts}

\myusepackage{units}%
\newcommand{\dr}{\nicefrac}

\myusepackage[nolabel]{fnbreak}%

\myusepackage{psfrag}%

\myusepackage[hyphens,spaces,obeyspaces]{url}%

\myusepackage{hyperref}%
\setpdflinkmargin{0.75pt}%

\myusepackage[usenames,dvipsnames]{xcolor}
\hypersetup{
colorlinks,
linkcolor={red!55!black},
citecolor={blue!55!black},
urlcolor={blue!35!black}
}

\myusepackage[linewidth=1pt]{mdframed}

\myusepackage{hyphenat}%

\myusepackage{caption}%

\myusepackage{sansmathfonts}%

\myusepackage{microtype}%

\myusepackage{marginnote}%

\myusepackage{calc}%

\myusepackage{mathrsfs}%

\myusepackage{enumitem}%

\myusepackage{graphicx}

\newcommand{\dgCapDefinition}{Definition}
\newcommand{\dgCapDefinitions}{Definitions}
\newcommand{\dgCapPostulate}{Postulate}
\newcommand{\dgCapPostulates}{Postulates}
\newcommand{\dgCapExample}{Example}

\newcommand{\dgCapFact}{Fact}
\newcommand{\dgCapFacts}{Facts}
\newcommand{\dgCapQuestion}{Question}
\newcommand{\dgCapQuestions}{Questions}
\newcommand{\dgCapLemma}{Lemma}
\newcommand{\dgCapLemmas}{Lemmas}
\newcommand{\dgCapNotation}{Notation}
\newcommand{\dgCapCorollary}{Corollary}
\newcommand{\dgCapCorollaries}{Corollaries}
\newcommand{\dgCapProposition}{Proposition}
\newcommand{\dgCapPropositions}{Propositions}
\newcommand{\dgCapClaim}{Claim}
\newcommand{\dgCapClaims}{Claims}
\newcommand{\dgCapTheorem}{Theorem}
\newcommand{\dgCapTheorems}{Theorems}
\newcommand{\dgCapProblem}{Problem}
\newcommand{\dgCapProblems}{Problems}
\newcommand{\dgCapRemark}{Remark}
\newcommand{\dgCapRemarks}{Remarks}
\newcommand{\dgCapConjecture}{Conjecture}
\newcommand{\dgCapConjectures}{Conjectures}
\newcommand{\dgCapResult}{Result}

\newcommand{\dgCapPart}{Part}
\newcommand{\dgCapParts}{Parts}
\newcommand{\dgCapChapter}{Chapter}
\newcommand{\dgCapChapters}{Chapters}
\newcommand{\dgCapSection}{Section}
\newcommand{\dgCapSections}{Sections}
\newcommand{\dgCapSubsection}{Subsection}
\newcommand{\dgCapSubsections}{Subsections}
\newcommand{\dgCapFigure}{Figure}
\newcommand{\dgCapFigures}{Figures}

\newcommand{\dgProofOf}{\proofname\ of}

\newcommand{\dgDefinition}{Definition}
\newcommand{\dgDefinitions}{Definitions}
\newcommand{\dgPostulate}{Postulate}
\newcommand{\dgPostulates}{Postulates}

\newcommand{\dgFact}{Fact}
\newcommand{\dgFacts}{Facts}
\newcommand{\dgQuestion}{Question}
\newcommand{\dgQuestions}{Questions}
\newcommand{\dgLemma}{Lemma}
\newcommand{\dgLemmas}{Lemmas}
\newcommand{\dgCorollary}{Corollary}
\newcommand{\dgCorollaries}{Corollaries}
\newcommand{\dgProposition}{Proposition}
\newcommand{\dgPropositions}{Propositions}
\newcommand{\dgClaim}{Claim}
\newcommand{\dgClaims}{Claims}
\newcommand{\dgTheorem}{Theorem}
\newcommand{\dgTheorems}{Theorems}
\newcommand{\dgProblem}{Problem}
\newcommand{\dgProblems}{Problems}
\newcommand{\dgRemark}{Remark}
\newcommand{\dgRemarks}{Remarks}
\newcommand{\dgConjecture}{Conjecture}
\newcommand{\dgConjectures}{Conjectures}

\newcommand{\dgPart}{Part}
\newcommand{\dgParts}{Parts}
\newcommand{\dgChapter}{Chapter}
\newcommand{\dgChapters}{Chapters}
\newcommand{\dgSection}{Section}
\newcommand{\dgSections}{Sections}
\newcommand{\dgSubsection}{Subsection}
\newcommand{\dgSubsections}{Subsections}
\newcommand{\dgFigure}{Figure}
\newcommand{\dgFigures}{Figures}

\newcommand{\dgFootnote}{Footnote}
\newcommand{\dgFootnotes}{Footnotes}

\ifndef{theorem}{}
\ifndef{theorem*}{\newtheorem*{theorem*}{\dgCapTheorem}}

\ifndef{lemma}{}
\ifndef{lemma*}{\newtheorem*{lemma*}{\dgCapLemma}}
\ifndef{corollary}{}
\ifndef{corollary*}{\newtheorem*{corollary*}{\dgCapCorollary}}
\ifndef{conjecture}{}
\ifndef{conjecture*}{\newtheorem*{conjecture*}{\dgCapConjecture}}
\ifndef{proposition}{}
\ifndef{proposition*}{\newtheorem*{proposition*}{\dgCapProposition}}
\ifndef{claim}{}
\ifndef{claim*}{\newtheorem*{claim*}{\dgCapClaim}}
\ifndef{result}{}
\ifndef{result*}{\newtheorem*{result*}{\dgCapResult}}
\ifndef{problem}{}
\ifndef{problem*}{\newtheorem*{problem*}{\dgCapProblem}}

\newtheoremstyle{mydefinition}%
{\topsep}{\topsep}%
{\slshape}%
{}%
{\bfseries}%
{.}%
{ }%
{}%

\newtheoremstyle{mynotation}%
{\topsep}{\topsep}%
{}%
{}%
{\bfseries\itshape}%
{.}%
{ }%
{}%

\newtheoremstyle{myremark}%
{\topsep}{\topsep}%
{\slshape}%
{}%
{\bfseries\itshape}%
{.}%
{ }%
{\thmname{#1}\thmnumber{~#2}}%

\newtheoremstyle{myexample}%
{\topsep}{\topsep}%
{\itshape}%
{}%
{\slshape}%
{.}%
{ }%
{\underline{\thmname{#1}\thmnumber{~#2}}}%

\newtheoremstyle{myclaims}%
{\topsep}{\topsep}%
{\slshape}%
{}%
{\bfseries\slshape}%
{.}%
{ }%
{\thmname{#1}\thmnumber{~#2}\thmnote{\textnormal{~(#3)}}}%

\ifndef{notation}{\theoremstyle{mynotation}}

{\theoremstyle{myremark}

\newtheorem*{myremark*}{\dgCapRemark}
}
{\theoremstyle{mydefinition}
\ifndef{postulate}{}
\ifndef{postulate*}{\newtheorem*{postulate*}{\dgCapPostulate}}
\ifndef{definition}{}
\ifndef{definition*}{\newtheorem*{definition*}{\dgCapDefinition}}
}
{\theoremstyle{myexample}
\ifndef{example}{}
\ifndef{example*}{\newtheorem*{example*}{\dgCapExample}}
}
{\theoremstyle{myclaims}

\newtheorem*{my_claim*}{\dgCapClaim}
\ifndef{fact}{}
\ifndef{fact*}{\newtheorem*{fact*}{\dgCapFact}}
\ifndef{question}{}
\ifndef{question*}{\newtheorem*{question*}{\dgCapQuestion}}
}

\newtheoremstyle{anystatementst}%
{\topsep}{\topsep}%
{\itshape}%
{}%
{\bfseries}%
{.}%
{ }%
{#3}%

{\theoremstyle{anystatementst} }

\newcommand{\newident}[3][\MyUniPat]{\ifthenelse{\equal{\MyUniPat}{#1}}
{
\newcommand{#2}[1][]{\ensuremath{\mathit{#3##1}}}
}
{\ifthenelse{\equal{}{#1}}
{
\newcommand{#2}[1][]{\ensuremath{\mathit{#3}}}
}
{
\DeclareRobustCommand{#2}[1][\MyUniPat]{\ifthenelse{\equal{\MyUniPat}{##1}}%
{\ensuremath{\mathit{#1}}}%
{\ensuremath{\mathit{#3}}}}
}
}
}

\newcommand{\newidenT}[3][\MyUniPat]{\ifthenelse{\equal{\MyUniPat}{#1}}
{
\DeclareRobustCommand{#2}[1][\MyUniPat]{\ifthenelse{\equal{\MyUniPat}{##1}}%
{\il{#3}}%
{\ensuremath{\mathit{#3##1}}}}
}
{
\DeclareRobustCommand{#2}[1][\MyUniPat]{\ifthenelse{\equal{\MyUniPat}{##1}}%
{\il{#1}}%
{\ensuremath{\mathit{#3}}}}
}
}

\newcommand{\newmat}[3][\MyUniPat]{\ifthenelse{\equal{\MyUniPat}{#1}}%
{\newcommand{#2}[1][]{#3##1}}%
{\newcommand{#2}[1][]{#3}}%
}

\newcommand{\providemat}[3][\MyUniPat]{\ifthenelse{\equal{\MyUniPat}{#1}}
{\providecommand{#2}[1][]{#3##1}}
{\providecommand{#2}[1][]{#3}}%
}

\newcommand{\newmatop}[3][\MyUniPat]{\ifthenelse{\equal{\MyUniPat}{#1}}
{
\newcommand{#2}{\operatorname{#3}}
}
{
\DeclareRobustCommand{#2}[1][\MyUniPat]{\ifthenelse{\equal{\MyUniPat}{##1}}%
{\operatorname{#1}}%
{\operatorname{#3}}}
}
}

\newcommand{\newmatoparg}[3][\MyUniPat]{\ifthenelse{\equal{\MyUniPat}{#1}}
{
\DeclareRobustCommand{#2}[1]{\ifthenelse{\equal{}{##1}}{\operatorname{#3}}{\operatorname{#3}\lf(##1\rt)}}
}
{
\DeclareRobustCommand{#2}[2][\MyUniPat]{\ifthenelse{\equal{\MyUniPat}{##1}}%
{\ifthenelse{\equal{}{##2}}{\operatorname{#1}}{\operatorname{#1}\lf(##2\rt)}}%
{\ifthenelse{\equal{}{##2}}{\operatorname{#3}}{\operatorname{#3}\lf(##2\rt)}}}
}
}

\newcommand{\newOlike}[2]{
\DeclareRobustCommand{#1}[2][\MyUniPat]{\ifthenelse{\equal{\MyUniPat}{##1}}%
{\ensuremath{\mathit{#2}\lf(##2\rt)}}%
{#2(##2)}%
}}

\makeatletter
\newcommand{\MyMakeTheoMacros}[3]{
\expandafter\newcommand\csname\expandafter\@gobble\string#2NostarNoname@DGaux\endcsname[2][]
{\ifthenelse{\equal{}{##1}}%
{\begin{#1}~##2 \end{#1}}%
{\begin{#1}\label{##1}~##2\end{#1}}%
}
\expandafter\newcommand\csname\expandafter\@gobble\string#2StarNoname@DGaux\endcsname[1]
{\begin{#1*}~##1 \end{#1*}}
\newcommand#2{\expandafter\@ifstar%
\expandafter{\csname\expandafter\@gobble\string#2StarNoname@DGaux\endcsname}%
{\csname\expandafter\@gobble\string#2NostarNoname@DGaux\endcsname}%
}

\expandafter\newcommand\csname\expandafter\@gobble\string#2NostarName@DGaux\endcsname[3][]
{\ifthenelse{\equal{}{##1}}%
{\begin{#1}[\e{##2}]~##3 \end{#1}}%
{\begin{#1}[\e{##2}]\label{##1}~##3\end{#1}}%
}
\expandafter\newcommand\csname\expandafter\@gobble\string#2StarName@DGaux\endcsname[2]
{\begin{#1*}[\e{##1}]~##2 \end{#1*}}
\newcommand#3{\expandafter\@ifstar%
\expandafter{\csname\expandafter\@gobble\string#2StarName@DGaux\endcsname}
{\csname\expandafter\@gobble\string#2NostarName@DGaux\endcsname}%
}
}

\newtheorem*{rep@theorem}{\rep@title}
\newcommand{\newreptheorem}[2]{%
\newenvironment{rep#1}[1]{%
\def\rep@title{#2 \ref{##1}}%
\begin{rep@theorem}}%
{\end{rep@theorem}}}
\makeatother

\newcommand{\MyMakeDupTheoMacros}[7]{
\MyMakeTheoMacros{#1}{#2}{#3}
\newreptheorem{#1}{#6}
\newcommand{#4}[3]{
\newcommand{##2}{##3}
\begin{#1}\label{##1}~##2\end{#1}}
\newcommand{#5}[4]{
\newcommand{##2}{##4}
\begin{#1}{\e{##3}}\label{##1}~##2\end{#1}}
\newcommand{#7}[2]{\begin{rep#1}{##1}~##2 \end{rep#1}}
}

\newcommand{\MyMakeRefMacros}[3]{\newcommand{#1}{\dgref{#2}{#3}}}

\MyMakeTheoMacros{fact}{\fct}{\nfct}

\MyMakeRefMacros{\fctref}{\dgFact~}{\dgFacts~}
\MyMakeRefMacros{\Fctref}{\dgCapFact~}{\dgCapFacts~}

\MyMakeTheoMacros{question}{\quest}{\nquest}

\MyMakeRefMacros{\questref}{\dgQuestion~}{\dgQuestions~}
\MyMakeRefMacros{\Questref}{\dgCapQuestion~}{\dgCapQuestions~}

\MyMakeTheoMacros{notation}{\nota}{\nnota}

\MyMakeDupTheoMacros{lemma}
{\lem}{\nlem}{\lemdup}{\nlemdup}{\dgLemma}{\lemrep}

\MyMakeRefMacros{\lemref}{\dgLemma~}{\dgLemmas~}
\MyMakeRefMacros{\Lemref}{\dgCapLemma~}{\dgCapLemmas~}

\MyMakeDupTheoMacros{corollary}
{\crl}{\ncrl}{\crldup}{\ncrldup}{\dgCorollary}{\crlrep}

\MyMakeRefMacros{\crlref}{\dgCorollary~}{\dgCorollaries~}
\MyMakeRefMacros{\Crlref}{\dgCapCorollary~}{\dgCapCorollaries~}

\MyMakeTheoMacros{proposition}{\prp}{\nprp}

\newtheorem*{prp*}{\e{\dgCapProposition}}

\MyMakeRefMacros{\prpref}{\dgProposition~}{\dgPropositions~}
\MyMakeRefMacros{\Prpref}{\dgCapProposition~}{\dgCapPropositions~}

\MyMakeDupTheoMacros{my_claim}
{\clm}{\nclm}{\clmdup}{\nclmdup}{\dgClaim}{\clmrep}

\MyMakeRefMacros{\clmref}{\dgClaim~}{\dgClaims~}
\MyMakeRefMacros{\Clmref}{\dgCapClaim~}{\dgCapClaims~}

\MyMakeDupTheoMacros{theorem}
{\theo}{\ntheo}{\theodup}{\ntheodup}{\dgTheorem}{\theorep}

\MyMakeRefMacros{\theoref}{\dgTheorem~}{\dgTheorems~}
\MyMakeRefMacros{\Theoref}{\dgCapTheorem~}{\dgCapTheorems~}

\MyMakeTheoMacros{postulate}{\postu}{\npostu}

\MyMakeRefMacros{\posturef}{\dgPostulate~}{\dgPostulates~}
\MyMakeRefMacros{\Posturef}{\dgCapPostulate~}{\dgCapPostulates~}

\MyMakeTheoMacros{definition}{\defi}{\ndefi}

\MyMakeRefMacros{\defiref}{\dgDefinition~}{\dgDefinitions~}
\MyMakeRefMacros{\Defiref}{\dgCapDefinition~}{\dgCapDefinitions~}

\MyMakeTheoMacros{problem}{\prob}{\nprob}

\MyMakeRefMacros{\probref}{\dgProblem~}{\dgProblems~}
\MyMakeRefMacros{\Probref}{\dgCapProblem~}{\dgCapProblems~}

\MyMakeTheoMacros{myremark}{\rem}{\nrem}

\MyMakeRefMacros{\remref}{\dgRemark~}{\dgRemarks~}
\MyMakeRefMacros{\Remref}{\dgCapRemark~}{\dgCapRemarks~}

\MyMakeTheoMacros{conjecture}{\conj}{\nconj}

\MyMakeRefMacros{\conjref}{\dgConjecture~}{\dgConjectures~}
\MyMakeRefMacros{\Conjref}{\dgCapConjecture~}{\dgCapConjectures~}

\renewcommand{\qedsymbol}{$\blacksquare$}

\newcommand{\prfstart}[1][]{\ifthenelse{\equal{}{#1}}%
{\begin{proof}\renewcommand{\qedsymbol}{$\blacksquare$}}%
{\begin{proof}[\dgProofOf\ #1]%
\renewcommand{\qedsymbol}{$\blacksquare_{\mbox{\it{\scriptsize{#1}}}}$}}%
}
\newcommand{\prfend}[1][*]{%
\ifthenelse{\equal{}{#1}}{\renewcommand{\qedsymbol}{$\blacksquare$}}{}%
\ifthenelse{\equal{*}{#1}}{}%
{\renewcommand{\qedsymbol}{$\blacksquare_{\mbox{\it{\scriptsize{#1}}}}$}}%
\end{proof}\renewcommand{\qedsymbol}{$\blacksquare$}%
}

\newcommand{\NewSectLikeDG}[3]{
\NewcommandThreeArgsTwoOpt{#1}
{\ifthenelse{\equal{*}{##2}}{#2*}{#2{##2}}}
{#2{##2\label{##1}}}
{#2[##3]{##2#3{##3}}#3{##3}}
{#2[##3]{##2#3{##3}\label{##1}}#3{##3}}
}

\NewSectLikeDG{\sect}{\section}{\sectionmark}

\NewSectLikeDG{\ssect}{\subsection}{\subsectionmark}

\NewSectLikeDG{\sssect}{\subsubsection}{\subsubsectionmark}

\makeatletter
\@addtoreset{chapter}{part}
\makeatother

\newcommand*\parttitle{}
\let\origpart\part
\renewcommand*{\part}[2][]{%
\ifx\\#1\\
\origpart{#2}%
\renewcommand*\parttitle{#2}%
\else
\origpart[#1]{#2}%
\renewcommand*\parttitle{#1}%
\fi
}

\ifndef{chapter}
{\NewSectLikeDG{\chap}{\part}{\DoNothing}}%
[\NewSectLikeDG{\chap}{\chapter}{\chaptermark}]

\NewSectLikeDG{\prt}{\part}{\DoNothing}%

\newcommand{\para}[2][]{\ifthenelse{\equal{}{#1}}
{\paragraph{#2}}
{\paragraph{#2\label{#1}}}}

\NewcommandTwoArgsLastOpt{\parass}
{\para{#1}\toc{subsection}{#1}}
{\para{#1#2}\toc{subsection}{#1}}

\makeatletter

\def\@seccntformat#1{\@ifundefined{#1@dgformat}%
{\csname the#1\endcsname\quad}%
{\csname #1@dgformat\endcsname.~}}%

\renewcommand\theparagraph{\arabic{paragraph}}
\newcommand{\paragraph@dgformat}{\S\theparagraph}

\renewcommand\thesubparagraph{\arabic{paragraph}.\arabic{subparagraph}}
\newcommand{\subparagraph@dgformat}{\S\thesubparagraph}

\MyMakeRefMacros{\paref}{\S}{\S\S~}
\makeatother

\MyMakeRefMacros{\prtref}{\dgPart~}{\dgParts~}
\MyMakeRefMacros{\Prtref}{\dgCapPart~}{\dgCapParts~}

\MyMakeRefMacros{\chref}{\dgChapter~}{\dgChapters~}
\MyMakeRefMacros{\Chref}{\dgCapChapter~}{\dgCapChapters~}

\MyMakeRefMacros{\sref}{\dgSection~}{\dgSections~}
\MyMakeRefMacros{\Sref}{\dgCapSection~}{\dgCapSections~}

\MyMakeRefMacros{\ssref}{\dgSubsection~}{\dgSubsections~}
\MyMakeRefMacros{\Ssref}{\dgCapSubsection~}{\dgCapSubsections~}

\MyMakeRefMacros{\sssref}{\dgSubsection~}{\dgSubsections~}
\MyMakeRefMacros{\Sssref}{\dgCapSubsection~}{\dgCapSubsections~}

\newcommand{\toc}[3][]{%
\ifthenelse{\equal{chapter}{#2}}{\newpage}{}%
\ifthenelse{\equal{part}{#2}}{\newpage}{}%
\ifthenelse{\equal{}{#1}}%
{\phantomsection\addcontentsline{toc}{#2}{#3}}%
{\phantomsection\refstepcounter{#2}\label{#1}\addcontentsline{toc}{#2}{#3}}%
}

\newcommand{\toct}[1]{\toc{section}{#1}}

\definecolor{DarkRed}{rgb}{0.65,0.05,0.05}

\definecolor{LightRed}{rgb}{0.855,0.16,0.16}

\definecolor{DarkGreen}{rgb}{0, 0.52, 0.05}

\definecolor{LightGreen}{rgb}{0.08,0.855,0.16}

\definecolor{DarkBlue}{rgb}{0.05, 0, 0.55}

\definecolor{LightBlue}{rgb}{0.122,0.016,0.855}

\MyMakeRefMacros{\figref}{\dgFigure~}{\dgFigures~}
\MyMakeRefMacros{\Figref}{\dgCapFigure~}{\dgCapFigures~}

\newcommand{\IfMathMode}[2]{\ifmmode{#1}\else{#2}\fi}

\newcommand{\fbr}[1]{%
\delimiterfactor=1001%
\IfMathMode%
{#1}{$#1$}\delimiterfactor=901%
}

\newcommand{\fnbr}[1]{\mbox{\fbr{#1}}}%

\newcommand{\f}[2][*]{\ifthenelse{\equal{}{#1}}{\fbr{#2}}{\fnbr{#2}}}

\newcommand{\mal}[2][]{\MyChangeMathMargins%
\delimiterfactor=1001%
\ifthenelse{\equal{}{#1}}%
{\begin{align*} #2 \end{align*}}%
{\ifthenelse{\equal{P}{#1}}%
{\allowdisplaybreaks\begin{align*} #2%
\end{align*}\interdisplaylinepenalty=10000}%
{\begin{align}\begin{split}\label{#1} #2 \end{split}\end{align}}%
}\delimiterfactor=901%
}

\newcommand{\m}{\mal}

\makeatletter

\DeclareRobustCommand\bref{\@ifnextchar[%
{\bref@presq}%
{\bref@nopresq}%
}

\def\bref@presq[#1]{\@ifnextchar[%
{(\ref{#1}), \bref@presq}
{(\ref{#1}) and~\bref@nopresq}
}

\def\bref@nopresq#1{(\ref{#1})}

\DeclareRobustCommand\dgref[2]{\@ifnextchar[%
{#2\dgref@presq}%
{#1\dgref@nopresq}%
}

\def\dgref@presq[#1]{\@ifnextchar[%
{\ref{#1}, \dgref@presq}
{\ref{#1} and~\dgref@nopresq}
}

\def\dgref@nopresq#1{\ref{#1}}

\newcommand\Cases[1][0pt]{%
\def\Case@skipam{#1}%
\left\{\!\!\!\begin{array}{ll}\Cases@continue%
}

\def\Cases@continue#1#2{\@ifnextchar\bgroup%
{#1 &\txt{#2}\\[\Case@skipam] \Cases@continue}%
{#1 &\txt{#2}\end{array}\Cases@end}%
}

\def\Cases@end{\@ifnextchar[%
{\Cases@end@postsq}%
{\right.}
}

\def\Cases@end@postsq[#1]{\ifthenelse{\equal{\}}{#1}}
{\!\!\right\}}%
{\right.}
}

\makeatother

\newcommand{\lf}{\mathopen{}\mathclose\bgroup\left}
\newcommand{\rt}{\aftergroup\egroup\right}

\providecommand{\middle}{\big}
\newcommand{\md}{\middle}

\newcommand{\ud}{\vphantom{|_1^1}}

\newcommand{\chs}{\genfrac(){0cm}{}}%

\newmatoparg{\poly}{poly}
\newmatoparg{\qpoly}{qpoly}
\newmatoparg{\plog}{poly-log}
\newmatoparg[disc]{\disc}{disc_{#1}}
\newmatoparg{\sign}{sign}
\newmatoparg{\rank}{rank}
\newmatoparg{\tr}{tr}
\newmatoparg{\spn}{span}

\newmatoparg{\supp}{supp}%
\newmatoparg{\sgn}{sgn}%
\newmatoparg{\diam}{diam}%
\newmatoparg{\Dim}{dim}%
\newmatoparg{\pow}{pow}

\newcommand{\NewHLikeDG}[2]{
\NewcommandThreeArgsTwoOpt{#1}
{\operatorname{\mathnormal{#2}}\lf(##2\rt)}
{\operatorname{\mathnormal{#2_{##1}}}\lf(##2\rt)}
{\operatorname{\mathnormal{#2}}\lf(##2\md|{\ud}##3\rt)}
{\operatorname{\mathnormal{#2_{##1}}}\lf(##2\md|{\ud}##3\rt)}
}

\NewHLikeDG{\h}{H}

\NewHLikeDG{\hm}{H_{min}}

\newcommand{\NewILikeDG}[2]{
\NewcommandFourArgsTwoOpt{#1}
{\mathop{\pmb{#2}}\lf[##2:{\ud}##3\rt]}
{\mathop{\pmb{#2}\?\!_{##1}}\lf[##2:{\ud}##3\rt]}
{\mathop{\pmb{#2}}\lf[##2:##3\md|{\ud}##4\rt]}
{\mathop{\pmb{#2}\?\!_{##1}}\lf[##2:##3\md|{\ud}##4\rt]}
}

\NewILikeDG{\I}{I}

\newcommand{\NewELikeDG}[2]{
\NewcommandThreeArgsTwoOpt{#1}
{#2\lf[##2\rt]}
{#2_{##1}\lf[##2\rt]}
{#2\lf[##2\md|{\ud}##3\rt]}
{#2_{##1}\lf[##2\md|{\ud}##3\rt]}
}

\NewELikeDG{\PR}{\mathop{\pmb{Pr}}}

\NewELikeDG{\E}{\mathop{\pmb{E}}}

\NewELikeDG{\Del}{\mathop{\pmb{\Delta}}}

\NewELikeDG{\Var}{\mathop{\pmb{Var}}}

\providecommand{\U}{}%
\renewcommand{\U}[1][]{\ifthenelse{\equal{}{#1}}%
{\Cl U}%
{\Cl U_{#1}}}

\NewcommandTwoArgsLastOpt{\GF}
{{\mathcal{G\!F}_{#1}}}
{{\mathcal{G\!F}_{#1}^{#2}}}

\providemat{\NN}{\mathbb{N}}
\providemat{\RR}{\mathbb{R}}

\newcommand*{\wht}{\widehat}

\newcommand*{\wtl}{\widetilde}

\makeatletter

\def\wbr@scale#1#2{%
\scalebox{#2}{
\vbox{%
\hrule height 0.78pt%
\kern0.15ex%
\hbox{%
\kern-0.1em%
\ensuremath{#1}%
\kern-0.05em%
}%
}%
}%
}

\newcommand*{\wbr}[1]{%
\ifmmode\mathchoice%
{\wbr@scale{#1}{1}}%
{\wbr@scale{#1}{1}}%
{\wbr@scale{#1}{0.75}}%
{\wbr@scale{#1}{0.58}}%
\else%
\wbr@scale{#1}{1}%
\fi%
}

\makeatother

\newcommand*{\wbC}[1]{\wbr{\Cl{#1}}}%

\newcommand{\pl}[1][]{\ifthenelse{\equal{}{#1}}%
{\mskip-6mu\stackrel{\text-}{}\mskip-4mu\txt{s}}%
{\f{#1\mskip-6mu\stackrel{\text-}{}\mskip-4mu\txt{s}}}}

\newcommand{\ord}[1][]{\ifthenelse{\equal{}{#1}}%
{\txt{'th}}%
{\ifthenelse{\equal{1}{#1}}{$1\txt{'st}$}{\ifthenelse{\equal{2}{#1}}{$2\txt{'nd}$}{\ifthenelse{\equal{3}{#1}}{$3\txt{'rd}$}{\f{#1\txt{'th}}}}}}}

\newcommand{\fr}[3][*]{%
\ifthenelse{\equal{*}{#1}}%
{\frac{#2}{#3}}{}%
\ifthenelse{\equal{}{#1}}%
{\dr{#2}{#3}}{}%
\ifthenelse{\equal{/}{#1}}%
{\lf.#2\md/#3\rt.}{}%
\ifthenelse{\equal{p_}{#1}}%
{\lf.\lf(#2\rt)\md/#3\rt.}{}%
\ifthenelse{\equal{_p}{#1}}%
{\lf.#2\md/\lf(#3\rt)\rt.}{}%
\ifthenelse{\equal{pp}{#1}}%
{\lf.\lf(#2\rt)\md/\lf(#3\rt)\rt.}{}%
}

\newcommand{\sq}{\sqrt}

\NewcommandThreeArgsTwoOpt{\set}
{\lf\{#2\rt\}}
{}
{\lf\{#2\md|\ud#3\rt\}}
{\lf\{#2\ud#1\ud#3\rt\}}

\newcommand{\NewMinLikeDG}[2]{
\NewcommandThreeArgsTwoOpt{#1}
{#2\lf\{##2\rt\}}
{#2_{##1}\lf\{##2\rt\}}
{#2\lf\{##2\md|\ud##3\rt\}}
{#2_{##1}\lf\{##2\md|\ud##3\rt\}}
}

\NewMinLikeDG{\Min}{\min}

\NewMinLikeDG{\Max}{\max}

\newmatop{\argmin}{argmin}
\NewMinLikeDG{\Argmin}{\argmin}

\newmatop{\argmax}{argmax}
\NewMinLikeDG{\Argmax}{\argmax}

\NewMinLikeDG{\Sup}{\sup}

\NewMinLikeDG{\Inf}{\inf}

\newOlike{\asO}{O}
\newOlike{\astO}{\wtl O}
\newOlike{\aso}{o}
\newOlike{\asOm}{\Omega}
\newOlike{\astOm}{\wtl\Omega}
\newOlike{\asom}{\omega}
\newOlike{\asT}{\Theta}
\newOlike{\astT}{\wtl\Theta}

\makeatletter

\DeclareRobustCommand\bk{\@ifnextchar[%
{\lf\langle \bk@bra}%
{\@ifnextchar<%
{\lf.\bk@op}
{\lf| \bk@ket}%
}%
}

\def\bk@bra[#1]{#1 \@ifnextchar[%
{\md|\hspace{-1.5pt}\md\langle \bk@bra}%
{\@ifnextchar\bgroup%
{\md| \bk@ket}%
{\@ifnextchar<%
{\md| \bk@op}%
{\rt|}%
}%
}%
}

\def\bk@ket#1{#1 \@ifnextchar[%
{\md\rangle\hspace{-1.5pt}\md\langle \bk@bra}%
{\@ifnextchar\bgroup%
{\md\rangle\hspace{-1.5pt}\md| \bk@ket}%
{\@ifnextchar<%
{\md\rangle \bk@op}%
{\rt\rangle}%
}%
}%
}

\def\bk@op<#1>{#1 \@ifnextchar[%
{\md\langle \bk@bra}%
{\@ifnextchar\bgroup%
{\md| \bk@ket}%
{\@ifnextchar<%
{\tm \bk@op}%
{\rt.}%
}%
}%
}

\makeatother

\providecommand{\ip}[2]{\lla #1\,,\,#2\rra}

\NewcommandTwoArgsLastOpt{\sz}
{\lf|#1\rt|}
{\lf|#1\rt|_{#2}}

\NewcommandTwoArgsLastOpt{\norm}
{\lf\|#1\rt\|}%
{\lf\|#1\rt\|_{#2}}

\providecommand{\floor}[2][*]{\ifthenelse{\equal{}{#1}}%
{\lfloor #2 \rfloor}%
{\llf #2 \rrf}}

\newcommand{\txt}[1]{\textrm{#1}}%

\newcommand{\Cl}{\mathcal}%

\DeclareMathAlphabet{\mathbfcal}{OMS}{cmsy}{b}{n}

\DeclareMathAlphabet{\mathlowcal}{OT1}{pzc}{m}{it}

\newidenT{\Pp}{P}
\newidenT[BPP]{\BPP}{BPP_{#1}}
\newidenT{\NP}{NP}

\newcommand{\lla}{\lf\langle}
\newcommand{\rra}{\rt\rangle}
\newcommand{\llf}{\lf\lfloor}
\newcommand{\rrf}{\rt\rfloor}

\newcommand{\nin}{\not\in}%

\newcommand{\dt}{\cdot}
\newcommand{\tm}{\cdot}
\newcommand{\deq}{\stackrel{\textrm{def}}{=}}
\newcommand{\smin}{\setminus}

\newcommand{\sbseq}{\subseteq}
\newcommand{\sbs}{\subset}

\newcommand{\eps}{\varepsilon}

\newcommand{\es}{\emptyset}

\newcommand{\biglor}{\bigvee}

\newcommand{\lxor}{\oplus}
\newcommand{\biglxor}{\bigoplus}

\NewcommandFourArgsTwoOptDefa{\overlay}
{0mu}{1}
{
\begingroup
\mathchoice{
\bgroup
\scalebox{#4}{\ooalign{$\displaystyle#2\mkern#1$\cr
\hidewidth{$\displaystyle\mkern#1#3$}\hidewidth}}
\egroup
}{
\bgroup
\scalebox{#4}{\ooalign{$\textstyle#2\mkern#1$\cr
\hidewidth{$\textstyle\mkern#1#3$}\hidewidth}}
\egroup
}{
\bgroup
\scalebox{#4}{\ooalign{$\scriptstyle#2\mkern#1$\cr
\hidewidth{$\scriptstyle\mkern#1#3$}\hidewidth}}
\egroup
}{
\bgroup
\scalebox{#4}{\ooalign{$\scriptscriptstyle#2\mkern#1$\cr
\hidewidth{$\scriptscriptstyle\mkern#1#3$}\hidewidth}}
\egroup
}
\endgroup
}

\newcommand{\unin}{\mathrel{\overlay[1.25mu]{\subset}{\sim}}}

\makeatletter
\def\mov@rlay#1#2{\leavevmode\vtop{%
\baselineskip\z@skip \lineskiplimit-\maxdimen
\ialign{\hfil$\m@th#1##$\hfil\cr#2\crcr}}}
\makeatother

\newcommand{\lcur}{\prec}
\newcommand{\lcure}{\preceq}

\newcommand{\ds}[1][]
{\ifthenelse{\equal{}{#1}}{\allowbreak\dots}{#1\allowbreak\dots#1}}
\newmat{\dc}{\ds[,]}
\newmat{\dpl}{\ds[+]}

\newmat{\OI}{\set{0,1}} 

\mathchardef\myhyphen="2D

\makeatletter

\let\dgampersand\&

\DeclareRobustCommand\&{%
\new@ifnextchar[%
{\dgsep@reposit}%
{\dgampersand}%
}

\def\dgsep@reposit[#1]{\hspace{#1}&\hspace{-#1}}

\makeatother

\newcommand{\abstart}{\begin{abstract}}
\newcommand{\abend}{\end{abstract}}

\newenvironment{myepig}
{\par\addtolength{\leftskip}{28mm}\addtolength{\rightskip}{8mm}\noindent\ignorespaces}
{\par}

\newenvironment{myepigsgn}
{\par\addtolength{\leftskip}{84mm}\noindent\ignorespaces}
{\par}

\NewcommandThreeArgsTwoOpt{\epig}
{{\small \begin{myepig} \textit{#2} \end{myepig} \addvspace{\baselineskip}}}
{\ClassError{My Macros}{The first optional argument is not expected in "\epig"}{}}
{{\small \begin{myepig} \textit{#2} \end{myepig} \Nopagebreak%
\begin{myepigsgn} -- #3 \end{myepigsgn}\addvspace{\baselineskip}}}
{\ClassError{My Macros}{The first optional argument is not expected in "\epig"}{}}

\newcommand{\cent}[1]{\begin{center} #1 \end{center}}

\newcommand{\itstart}[1][\MyUniPat]{%
\ifthenelse{\equal{\MyUniPat}{#1}}%
{\begin{itemize}[noitemsep,topsep=3pt]}%
{\ifthenelse{\equal{*}{#1}}%
{\begin{itemize}[noitemsep,topsep=3pt,leftmargin=*]}%
{\begin{itemize}[#1]}%
}%
}

\newcommand{\itend}{\end{itemize}}

\newcommand{\DoNothing}[1]{}%

\makeatletter

\protected \def \dg{\@ifstar\dg@st\dg@nost}%

\protected \def \dg@nost #1{%
\textcolor{Red}
{
{\normalmarginpar\marginnote{\bl{DG's note}}}
{\reversemarginpar\marginnote{\bl{DG's note}}\\}
\IfMathMode{
~~~\txt{#1}~
}{
~\\~~~#1~\\
{\normalmarginpar\marginnote{\bl{\ul{------}}}}
{\reversemarginpar\marginnote{\bl{\ul{------}}}\\}
}
}

}

\protected \def \dg@st #1{%
~\\\textcolor{Red}{%
$\blacktriangleleft\blacktriangleright$\\\\
#1\\\\
$\blacktriangleleft\blacktriangleright$}%
}

\makeatother

\newcommand{\fn}[2][]{%
\IfMathMode{}{}%
\ifthenelse{\equal{}{#1}}%
{\footnote{
\ignorespaces #2\unskip}}%
{\footnote{\label{#1}
\ignorespaces #2}}%
}

\newcommand{\fnm}{\footnotemark}
\newcommand{\fnt}[2][]{\ifthenelse{\equal{}{#1}}%
{\footnotetext{
\ignorespaces #2}\noindent\ignorespaces}%
{\footnotetext{\label{#1}
\ignorespaces #2}\noindent\ignorespaces}%
}

\MyMakeRefMacros{\fnref}{\dgFootnote~}{\dgFootnotes~}

\makeatletter

\newcommand{\Nopagebreak}{\par\nobreak\@afterheading} 

\makeatother

\DeclareTextFontCommand{\bemph}{\bfseries}
\DeclareTextFontCommand{\ibemph}{\bfseries\em}
\DeclareTextFontCommand{\ttl}{\ttfamily}

\newcommand{\e}{\emph}
\newcommand{\eb}{\bemph}

\newcommand{\eu}[1]{\ttl{\ul{#1}}}
\newcommand{\eui}[1]{\emph{\ul{#1}}}

\newcommand{\bl}[1]{{\bf #1}}%

\newcommand{\il}[1]{{\it #1}}%

\def\?{\mskip 0.5\thinmuskip}%

\newcounter{qdcount}
\newcommand{\qd}[1][1]{%
\setcounter{qdcount}{#1}%
\whiledo{\value{qdcount}>0}{\quad\addtocounter{qdcount}{-1}}%
}

\newcommand{\vsp}[1][1]{{\vskip #1\smallskipamount\vskip 0pt}}

\newcommand{\MyChangeMathMargins}{%
\setlength{\abovedisplayskip}{\abovedisplayshortskip + 4pt}%
\setlength{\belowdisplayskip}{\abovedisplayshortskip + 5pt}%
}

\newident{\Rq}{R^q}
\newident{\bRq}{\wbr{R^q}}

\newident{\Dq}{D^q}

\newident{\Rx}{R^\lxor}
\newident{\bRx}{\wbr{R^\lxor}}

\newident{\Dx}{D^\lxor}

\newident{\Sea}{Search}

\newident{\ApMa}{\wtl{Maj}}

\newident{\muCt}{\mu_C^{(t)}}

\title{Unambiguous parity-query complexity\fn
{This is a preliminary version...}}

\newcommand{\instDG}{Institute of Mathematics of the Czech Academy of Sciences, \v Zitna 25, Praha 1, Czech Republic.}

\newcommand{\DmytroG}{In 2022 the author changed his first name, as explained on his Internet page.}

\newcommand{\thanksDG}{Partially funded by the grant 19-27871X of GA \v CR and by RVO:\ 67985840.}

\author{Dmytro Gavinsky\thanks{\DmytroG} \thanks{\instDG\newline\thanksDG}
}

\begin{document}

\maketitle

\thispagestyle{empty}

\abstart

We give a lower bound of \asOm{\sq n} on the \e{unambiguous randomised parity-query complexity} of the \e{approximate majority} problem -- that is, on the lowest randomised parity-query complexity of any \e{function} over $\OI^n$ whose value is ``$0$'' if the Hamming weight of the input is at most $\dr n3$, is ``$1$'' if the weight is at least $\dr {2n}3$, and may be arbitrary otherwise.

\abend

\sect[s_intro]{Introduction}

The computational model of \e{parity queries} is a well-known natural strengthening of the (more common) bit-wise query model:\ in both models the input is $x\in \OI^n$, in the bit-query model a protocol can get the value of a single bit of $x$ at each step, while in the parity-query model it can, for any $s\sbseq\set{1\dc n}$, obtain the value of $\biglxor_{i\in s}x_i$ via a single query.
Alternatively, one may view a parity-query protocol as receiving its input $x$ from the binary linear space $\GF2[n]$ and being able to make an arbitrary linear query to the coordinates of $x$ at each step.

From the combinatorial standpoint, in the standard bit-query model a deterministic protocol partitions its input space $\OI^n$ into same-answer \e{sub-cubes}, while a deterministic parity-query protocol partitions its input space $\GF2[n] \simeq \OI^n$ into same-answer \e{affine subspaces}, and so, the parity-query model is a linear closure of the bit-query one.
The corresponding randomised models are the convex closures of their deterministic counterparts.
The parity-query model is, obviously, at least as strong as the bit-query one, and it is not hard to see that it can be much stronger (e.g., both deterministic and randomised bit-query complexity of computing $x_1\lxor x_2\ds[\lxor] x_n$ is $n$).

Let $g$ be a \e{relational computational problem} -- that is, one that admits non-unique correct answers for some input values.
Let $\Cl C$ be a complexity measure that is applicable to $g$.
We will denote by $\wbC C$ the \e{unambiguous-$\Cl C$} -- that is, let $\wbC C(g)$ be the minimum of $\Cl C(f_g)$ over all \e{functions} $f_g$ that have the same domain as $g$ and values that are consistent with $g$ (that is, $f_g(x)$ is a correct answer to $g(x)$ for every $x$ in the domain).

Our notion of \e{unambiguous algorithms} (or protocols) was studied in other computational regimes under the names of \e{Bellagio algorithms} and \e{pseudo-deterministic algorithms}.
Italophile himself, the author would happily accept the former term, but unfortunately it is not commonly used and the connotation might not be obvious.
As for the latter, it perfectly fits models like bit-wise query complexity, where randomised algorithms cannot outperform qualitatively the deterministic ones in computing functional problems:\ in that case forcing unambiguous answers -- that is, turning the original relational problem into a functional one -- means depriving the computing algorithms of its ``randomised privileges''.
On the other hand, for structurally richer settings like parity queries or communication complexity, the requirement of unambiguity is unrelated to the protocol's being deterministic, as the qualitative superiority of randomness in those models can be manifested in solving functional problems as well.
We will use the term \e{unambiguity} with respect to all computational settings considered in this work.

The concept of unambiguity highlights some alluring gaps in the current understanding of randomised computation.
For instance, consider the \e{approximate majority} relation -- one of the simplest examples that demonstrate the ``power of randomness'' -- here the answer must be ``$0$'' if the Hamming weight of the input $x\in\OI^n$ is at most $\dr n3$, ``$1$'' if the weight is at least $\dr {2n}3$ and may be arbitrary otherwise.
\itstart
\item In the standard bit-query model both the deterministic (\asOm n) and the randomised (\asO1) complexities of approximate majority can be established via nearly trivial arguments;
\item the unambiguous randomised query complexity (\asOm n) demands a somewhat more involved proof;
\item if we strengthen the model slightly by admitting parity queries, then bounding the unambiguous randomised complexity becomes even more challenging (as witnessed by this work), while both the deterministic and the randomised cases remain trivial;
\item similar situation can be expected in other ``rich enough'' models, like communication complexity.\fn
{
To the best of our knowledge, no non-trivial lower bound on the unambiguous complexity is known in communication complexity and this paper presents the first such example in the parity-query model.
}
\itend

The intricacy of analysing the unambiguous complexity can be reflected by the ``added universal quantifier'' in the corresponding formal statement.
Say, a usual claim of intractability of a relational task $g$ can be interpreted as
\m[intro_1]{
\txt{for \e{\bl{any} efficient protocol $\Pi$ there \ul{exists} input $x$, such that $\Pi(x)$ disagrees with $g(x)$}}
,}
while a claim of its \e{unambiguous} intractability would stand for
\m[intro_2]{
&\txt{\e{for \eb{any} function $f_g$ that is consistent with $g$}}\\
&\txt{and \e{\eb{any} efficient protocol $\Pi$ there \ul{exists} input $x$, such that $\Pi(x) \ne f_g(x)$}}
.}
The former statement implies the latter, which is not surprising:\ if $g$ admits no efficient randomised protocol, then it admits no efficient unambiguous randomised protocol either.
If, on the other hand, a relational problem does have an efficient randomised protocol -- which is the case, in particular, for approximate majority in the parity-query model -- then arguing its unambiguous randomised intractability requires \e{distinguishing} between statements \bref{intro_1} and \bref{intro_2} in the analysis, which may require rather fine tuning of the argument.

The state of affairs when one seems to understand well \e{why approximate majority is hard for deterministic protocols}, as well as \e{why the problem is easy for randomised protocols}, but not \e{why any function that computes approximate majority is hard for randomised protocols} seems to imply that one does not understand well the role of randomness in computing approximate majority -- one of the ``most canonical'' problems demonstrating the utility of randomness in query protocols.

\ssect[ss_tech]{Technical challenges}

And so, for this work we picked the setting of \e{parity queries} -- a model where determining \e{the unambiguous randomised complexity of approximate majority} looks challenging, but not hopeless.
This choice can be further justified via observing that three approaches that may look very natural at first glance are inherently incapable of providing a non-trivial lower bound on the quantity that we are interested in.
The first two of them are ``tactical'' approaches that work in the case of the standard bit-query model.

The \eui{most straightforward way} of arguing the hardness of unambiguity in the \e{bit-query model} is to apply the result of \cite{N91_CREW} stating that the randomised and the deterministic complexities of a function in that model are qualitatively the same (they can differ at most polynomially).
As the deterministic complexity of approximate majority is high, every function consistent with this relation is hard for deterministic bit-query protocols, and therefore for the randomised ones also.
Obviously, this approach is unsuitable for parity queries, as here the gap between the randomised and the deterministic complexity of a function can be as wide as \asO1\ vs.\ \asOm n\ (cf.~\clmref{c_Or}).

The \eui{second approach} that works well for the same case of the standard query model is even ``more ad hoc'', but also quantitatively tighter than the first one (it leads to the optimal bound of \asOm n, while the previous argument can only give \asOm{n^{\dr 13}}).
Let $f$ be a function consistent with \ApMa\ and computed by a randomised bit-query protocol, and let $x\in\OI^n$ be a point of (globally) maximal Hamming weight for which $f(x)=0$.
Such $x\in\OI^n$ must have Hamming weight less than $\dr{2n}3$, that is, $s\deq \set{i}[x_i=0]$ contains at least $\dr n3$ elements.
By the assumption, if $x'$ differs from $x$ only at some non-empty subset of bits whose indices are in $s$, then $f(x')=1$ -- that is, a protocol for $f$ makes enough queries, at least in some cases, in order to solve the \e{or} function on $|s|$ bits, and the randomised bit-query complexity of that task is, trivially, in $\asOm{|s|} = \asOm n$.
This argument cannot handle the case of parity-queries, as here the randomised complexity of the \e{or} function is in \asO1\ (again, cf.~\clmref{c_Or}).

The \eui{third approach} is one of the most efficient and widely applicable known strategies for analysing randomised complexity and, as such, is entitled to a section of its own.

\ssect[ss_minimax]{The minimax principle}

To prove a lower bound on the \e{randomised} complexity of certain task, one can start by guessing a ``hard'' input distribution $\mu$ and then prove that any \e{deterministic} protocol errs very often with respect to $\mu$.
On the one hand, as long as the family of randomised solutions in the considered computational model is the convex closure of the family of deterministic ones (which is the case not for all, but for most of natural computational models, in particular, for all those of interest to us in this work), Von Neumann's minimax principle~\cite{N28_Zur} implies that such hard distribution necessarily exists, so this approach can be viewed as universal.\fn
{
There are interesting models of computation where the randomised regime is not closed with respect to convex combinations of protocols:\ e.g., such is the case for the communication-complexity model of \e{simultaneous message passing with private randomness}, where the minimax principle in the above form doesn't hold (\e{equality} with constant error is hard in the worst-case regime but easy deterministically with respect to any fixed distribution).
}
On the other hand, the simpler structure of deterministic protocols usually makes their computational potential much easier to understand and to analyse than that of their ``convex generalisation'':\ the randomised protocols.

The regime of unambiguous complexity, on the other hand, is certainly not a ``mere convex combinations'' of deterministic protocols.
Recall that a relation $g$ has an efficient unambiguous solution if \e{there exists} some function $f$ that both agrees with $g$ and has an efficient solution.
Suppose that $g$ is hard, then \e{any $f$ that agrees with $g$ is hard} too, that is -- here applies the minimax principle principle in the above form -- \e{for any $f$ that agrees with $g$ there is a hard distribution $\mu_f$}.
These $\mu_f$, however, \e{may} be different for different functions $f$, not corresponding to any ``universally hard'' distribution for the relation $g$ itself.

What is more, these $\mu_f$ not only \e{can} but actually \e{must} disagree with one another, allowing for no single hard distribution for the relation $g$ -- at least, in all the qualitatively interesting cases, that is, as long as the gap between the deterministic and the randomised complexities of $g$ is significant.

Consider, for instance, a ``structurally meaningful'' case where $\Rx(g)$ is at most poly-logarithmic but both $\Dx(g)$ and $\bRx(g)$ are beyond that.
Then for any $\eps>0$ and distribution $\mu$ there exists a \e{deterministic} protocol $\Pi_{\mu,\eps}$ of complexity in \asO{\Rx(g)/\log \eps} that solves $g$ with error at most $\eps$ with respect to $\mu$ -- denote by $f_{\mu,\eps}$ the actual function computed by $\Pi_{\mu,\eps}$ (it is well-defined, as $\Pi_{\mu,\eps}$ is deterministic).
This $f_{\mu,\eps}$ disagrees with the relation $g$ with probability at most $\eps$ with respect to $\mu$.
Let $f_{\mu,\eps}'(\dt)$ be an arbitrary function with value $f_{\mu,\eps}(x)$ when $(x,\, f_{\mu,\eps}(x))\in g$ and some value from $\set{a}[(x,a)\in g]$ otherwise.
Note that $f_{\mu,\eps}'$ perfectly agrees with $g$ and $\Dx[_{\mu,\eps}](f_{\mu,\eps}') \in \asO{\Rx(g)/\log \eps}$, as witnessed by the protocol $\Pi_{\mu,\eps}$ (here ``\Dx[_{\mu,\eps}]'' stands for the deterministic complexity of solving the task with error at most $\eps$ with respect to $\mu$).

In other words, for every $\mu$ and $\eps>0$, there \e{always is} a function $f_{\mu,\eps}'$ that perfectly agrees with $g$ and for which there exists an efficient deterministic protocol with error at most $\eps$ with respect to $\mu$ -- in spite of the assumed intractability of the relation $g$ itself for \e{unambiguous} randomised protocols.\fn
{
The above argument readily adapts to virtually every reasonable model of computation with natural notions of determinism, randomness and unambiguity.
}
Therefore, the concept of unambiguous complexity can be viewed not only as ``computational randomness \e{without} the minimax principle'' (which by itself would likely be very interesting, although not unique), but as ``computational randomness \e{against} the minimax principle'' since the setting guarantees that either the considered case is structurally trivial or for every distribution $\mu$ the problem is easy.

\ssect*{This work}

To illustrate the structural richness of unambiguity, we present a lower bound of \asOm{\sq n} on the unambiguous randomised parity-query complexity of the approximate majority problem.
It follows that with respect to parity queries there exist:
\itstart
\item a problem that is intractable deterministically but easy for protocols with randomness, even under the requirement of unambiguity;
\item a problem that is intractable deterministically, easy with randomness but becomes intractable for randomised protocols if the requirement of unambiguity is imposed.
\itend

\ssect*{Prior work}

The concept of unambiguous complexity was introduced under the name of \e{Bellagio algorithms} by Gat and Goldwasser~\cite{GG11_Pro} and first studied in the context of the bit-wise query model under the name of \e{pseudo-determinism} by Goldreich, Goldwasser and Ron~\cite{GGR13_On}.

\sect[s_defi]{Preliminaries and definitions}[Preliminaries]

By default the logarithms are base-$2$.
We will write $[n]$ to denote the set $\set{1\dc n}\sbs\NN$ for $n\in \NN\cup \set{0}$ and let $[a] \deq [\Min{0,\floor a}]$ for $a\in \RR$.
Let $(a,\, b)$, $[a,\, b]$, $[a,\, b)$ and $(a,\, b]$ denote the corresponding open, closed and half-open intervals in $\RR$.
Towards readability, we will allow both $\set{\dt}[\dt]$ and $\set[:]{\dt}[\dt]$ to denote sets with conditions (preferring the former).
Let $\bot$ and $\top$ denote, respectively, the false and the true values.

For a linear space $S$, we will write ``$\le S$'', ``$< S$'', ``$\lcure S$'' and ``$\lcur S$'' to denote, respectively, its subspaces, proper subspaces, affine subspaces and proper affine subspaces:
\m{
A + b \lcure S
}
for any $A\le S$ and $b\in S$ (and similarly for ``$\lcur$'' and ``$<$'').
The zero element will be denoted by $\bar0$.

For any set $S$, we will denote by $\pow S$ the family of all its subsets and by $\chs St$ the family of its size-$t$ subsets.
We will write $\U[S]$ to denote the uniform distribution over the elements of $S$ and use the notation ``$X\sim\U[S]$'' and ``$X\unin S$'' interchangeably.
For a finite $S\sbs\NN$, we will write $S(i)$ to address the \ord[i] element of $S$ in natural ordering.

For $x\in\OI^n$, we let $\sz x$ denote its Hamming weight.
For $i\in{[n]}$, we will write either $x_i$ or $x(i)$ to address the \ord[i] bit of $x$ (preferring ``$x_i$'' unless it causes ambiguity) and for any $s\sbseq [n]$ both $x_s$ and $x(s)$ will stand for $x_{s(1)}\ds x_{s(|s|)}\in \OI^{|s|}$.

At times we will
assume the following trivial isomorphism:
\itstart
\item between the bit-strings $x\in \OI^n$ and the subsets $\set{i}[x_i=1]\sbseq [n]$ (in particular, the notation $\chs{[n]}k$ will stand for $\set{x\in\OI^n}[\sz{x}=k]$, and $x\cap y$ will address the set $\set{i\in[n]}[x_i=y_i=1]$);
\item between the \f n-bit strings and the elements of $\GF2[n]$ (in particular, the answer to the \e{linear} query represented by $s\sbseq[n]$ will be the parity $\biglxor_{i\in s}x_i$ for $x\in \OI^n$).
\itend
For $i\in [n]$, we will denote by $e_i$ both the unit vector in $\GF2[n]$ and the weight-$1$ bit string in $\OI^n$ that correspond to $\set{i}$.

We will use the following \eu{notation for affine subspaces of $\GF2[n]$}:
If $C\lcure\GF2[n]$, then $C'\le \GF2[n]$ is the ``supporting'' subspace for $C$, defined as
\m{
C' \deq C + C = C + c_0
}
for any $c_0\in C$ and $\wbr C\le \GF2[n]$ is the \e{dual} subspace (sometimes called the \e{annihilator}) of either $C$ or $C'$, defined as
\m{
\wbr C = \wbr{C'}
~\deq~ &\set{x\in \GF2[n]}[\forall\: y\in C':\: \ip xy=0]\\
=~ &\set{x\in \GF2[n]}[\exists\: c_x\in \GF2:\: \forall\: y\in C:\: \ip xy=c_x]
,}
where ``$\ip\dt\dt$'' stands for the inner product in $\GF2[n]$.~\fn[fn_duals]
{
As $\GF2[n]$ is not an inner product space for $n\ge 2$, the intersection of a linear subspace with its own dual can be a non-trivial linear subspace.
On the other hand, $\wbr{\wbr A} = A$ and $\Dim{A} + \Dim{\wbr A} = n$ for any linear subspace $A\sbseq \GF2[n]$.
}
The \e{co-dimension} of $C$ equals $n-\log\sz{C} = \Dim{\wbr C}$.

\ssect[ss_QuC]{Query complexity and unambiguity}[Queries and unambiguity]

The standard model of \e{query complexity} is among the simplest and the most natural settings for analysing the computational complexity of a Boolean function.

\ndefi[d_RDq]{\Rq\ and \Dq, (standard) query complexity}
{
Let $x\in \OI^n$ and $\Pi$ be a deterministic protocol that queries individual bits of $x$ and outputs a value denoted by $\Pi(x)$.
The \e{complexity} of $\Pi$ is the maximum possible number of queries that it makes.

A \e{randomised} query protocol is a convex combination of deterministic protocols:\ $\lf(\Pi_i, \alpha_i\rt)_{i}$ with $\sum_{i}\alpha_i=1$ outputs $\Pi_i(x)$ with probability $\alpha_i$ for every $i$ and $x$.
The complexity of such protocol is the maximum complexity of an individual $\Pi_i$.

The deterministic (randomised) query \e{complexity} of a function $f$, denoted by $\Dq(f)$ ($\Rq(f)$), is at most the complexity of a deterministic (randomised) query protocol that outputs $f(x)$ (with probability at least $\dr23$) for every input value $x$.

A query protocol is called \e{efficient} if its complexity is at most $\plog(n)$.
}

Obviously, every \Dq-protocol of complexity $k$ partitions the input space $\OI^n$ into at most $2^k$ monochromatic (with respect to the computed function) \e{sub-cubes} of co-dimension at most $k$.

The setting of \e{parity query complexity} is a natural strengthening of the standard model.

\ndefi[d_RDx]{\Rx\ and \Dx, parity query complexity}
{
Let $x\in \GF2[n]$ and $\Pi$ be a deterministic protocol that makes linear queries (or parity queries) to the bits of $x$, that is, for the query represented by $s\sbseq[n]$ the protocol receives the response $\biglxor_{i\in s}x_i$.
Denote the output of the protocol by $\Pi(x)$.
The \e{complexity} of $\Pi$ is the maximum possible number of linear queries that it makes.

A \e{randomised} parity-query protocol is a convex combination of deterministic protocols (cf.~\defiref{d_RDq}).

The deterministic (randomised) parity-query \e{complexity} of a function $f$, denoted by $\Dx(f)$ ($\Rx(f)$), is at most the complexity of a deterministic (randomised) parity-query protocol that outputs $f(x)$ (with probability at least $\dr23$) for every input value $x$.

A parity-query protocol is called \e{efficient} if its complexity is at most $\plog(n)$.
}

We will see in \sref{s_aff_GF2} that every \Dx-protocol of complexity $k$ partitions the input space $\GF2[n]$ into at most $2^k$ monochromatic (with respect to the computed function) \e{affine subspaces} of co-dimension at most $k$.

The primary context of this work is \e{structural complexity} and we will ask whether one computational setting can ``qualitatively outperform'' the other, that is, whether there is a computational problem that has an efficient solution in the model $\Cl M_1$, thought not in $\Cl M_2$.
In other words, computational problems are tools for separating computational models, and there is a class of problems that generalises the class of functions and, in some cases, give rise to model separations that wouldn't be possible via functions alone.

\ndefi[d_Rel]{relational problems}
{
Let $X$ be the domain, that is, the set of possible input values to a computational problem, and let $A$ be the range, that is, the set of possible answers.
Then a relation $g\sbseq X\times A$ defines the following computational problem:\ ``$a$'' is a \e{correct} answer with respect to the input value $x\in X$ if $(x,a)\in g$ and it is \e{wrong} otherwise.

A relation is called \e{partial} if for some input values there is no correct answer, that is, $\exists\: x\in X:\: \forall\: a\in A:\: (x,a)\nin g$; otherwise the relation is \e{total}.

A function $f:\: X\to A$ is said to be \e{consistent} with $g$ if all its answers are consistent with those of the relation, that is, $\forall\: x\in X:\: (x,\, f(x))\in g$.
}

In other words, relations admit ambiguous answers or no answer at all for some input values, while the functional problems are a special case with exactly one correct answer being assigned to every $x\in X$.
All relations considered in this work are \e{total}.\fn
{
Partial relations are usually interpreted as ``guarantees'' that only the input values for which there is a correct answer are to be expected:\ otherwise a protocol is allowed to answer ``anything''.
This is useful in the context of \e{partial functions} (or \e{promise functions}): these are relational problems with \e{at most} one correct answer corresponding to every $x\in X$ (as a class of computational problems, it is intermediate between functions and relations).
On the other hand, in the case of relational problems one may consider, instead of a partial relation $g$, the total one $g'\deq g \cup \set{x}[\forall\: a\in A:\: (x,a)\nin g]\times A$, as $g$ and $g'$ are describing the same computational problem.
}

In \defiref[d_RDq]{d_RDx} we have addressed the complexity of functional problems only.
There are at least two natural ways to generalise it for a relation $g\sbseq X\times A$:
\itstart
\item the complexity of a $g$ can be defined as the smallest complexity of a protocol that produces answers that are \e{correct} with respect to $g$;
\item alternatively, it can be defined as the smallest complexity of a function $f$ that has the same domain as $g$ and ``agrees'' with it answer-wise, that is, $\forall\: x\in X:\: (x, f(x))\in g$.
\itend
The corresponding pair of definitions are equivalent for the deterministic models \Dq\ and \Dx; on the other hand, for both \Rq\ and \Rx\ the first version is the standard notion of relational complexity, while the second one is the unambiguous complexity.

\ndefi[d_UnaD]{deterministic complexity of relations}
{
Let $g\sbseq X\times A$ be a relational problem and $\Cl C$ be either \Dq\ or \Dx.

The \e{$\Cl C$-complexity} of $g$, denoted by $\Cl C(g)$, is at most $\Cl C(f)$ for any $f:\: X\to A$ such that $\forall\: x\in X:\: (x, f(x))\in g$.
}

\ndefi[d_UnaR]{randomised complexity of relations; unambiguity}
{
Let $g\sbseq X\times A$ be a relational problem and $\Cl C$ be either \Rq\ or \Rx.

The \e{$\Cl C$-complexity} of $g$, denoted by $\Cl C(g)$, is at most the $\Cl C$-complexity of a randomised protocol from the corresponding query model that outputs with probability at least $\dr23$ a value from $\set{a\in A}[(x,a)\in g]$ for every $x\in X$.

The \e{unambiguous $\Cl C$-complexity}, denoted by $\wbC C(g)$, is at most $\Cl C(f)$ for any $f:\: X\to A$ such that $\forall\: x\in X:\: (x, f(x))\in g$.
}

\ssect[ss_Tasks]{Tasks to consider}

It follows readily from the definitions that
\m[m_str]{
&\Dq(g) \ge \Dx(g),
\Rq(g) \ge \Rx(g)
~~\txt{and}~~
\bRq(g) \ge \bRx(g);\\
&\Rq(g) \le \bRq(g) \le \Dq(g)
~~\txt{and}~~
\Rx(g) \le \bRx(g) \le \Dx(g)
}
for any relation $g$.
The inequalities in the first line of \bref{m_str} can correspond to separations of \asO1\ vs. \asOm n:\ this is witnessed, in particular, by the parity function $\biglxor_{i\in [n]}x_i$.

The first inequality chain of the second line of \bref{m_str} can correspond to at most polynomial gaps:\ it is known~\cite{N91_CREW} that $\Dq(g) \in \asO{(\Rq(g))^3}$, and therefore from the standpoint of structural complexity all the considered regimes of (standard) query complexity are equivalent.
This leaves us with the second chain, namely
\f{
\Rx(g) \le \bRx(g) \le \Dx(g)
.}
As we are going to study the effect of the unambiguity upon the \e{efficient computability} of tasks, we shall only consider those relations $g$ for which $\Rx(g) \in \plog(n)$ and $\Dx(g)$ is much higher, preferably in $n^{\asOm1}$.

Our example of $g_1$ such that $\Rx(g_1) = \bRx(g_1) \ll \Dx(g_1)$ is fairly simple:\ it is the \e{or} function.

\clm[c_Or]
{
Let $X=\OI^n$ and $g_1:\: X\to \OI$ be the \e{or} function:
\m{
g_1(x) = \lor(x) \deq \biglor_{i\in [n]}x_i
.}

Then
\m{
\Rx(g_1) = \bRx(g_1) \in \asO1
~~~\txt{and}~~~
\Dx(g_1) = n
.}
}

\prfstart

As $g_1$ is a function, $\Rx(g_1) = \bRx(g_1)$.

To compute $\lor(x)$, a protocol must check whether $x=0^n$ and output ``$0$'' if that is the case and ``$1$'' otherwise.

If $x\ne 0^n$, then more than half of non-empty $s\sbseq[n]$ satisfy $\biglxor_{i\in s}x_i = 1$ (let $x_{i_0}=1$, then $\biglxor_{i\in s'}x_i \ne \biglxor_{i\in s'\cup\set{i_0}}x_i$ for each $s'\sbseq[n]\smin \set{i_0}$ and $\biglxor_{i\in\es}x_i=0$).
Accordingly, a single parity-query with uniformly random non-empty $s\sbseq[n]$ will return ``$1$'' with probability greater than $\dr 12$ if $x\ne 0^n$ and ``$0$'' with certainty if $x=0^n$.
A constant number of independent random parity-queries allows solving $\lor(x)$ with arbitrarily small constant error probability.

On the other hand, if a \e{deterministic} protocol that has asked at most $n-1$ parity-queries answers ``$0$'', then, obviously, there is some $x\ne 0^n$ that is consistent with all the answers received protocol; accordingly, this protocol does not compute $\lor(x)$.

\prfend[\clmref{c_Or}]

A very similar behaviour with respect to unambiguity is demonstrated by the \e{search} relation, which can be viewed as a natural relational version of the function $\lor(x)$.

\crl[cr_Sea]
{
Let $X=\OI^n$ and $\Sea\sbseq X\times [n]$ be the \e{search} relation:
\m{
g_1' = \Sea \deq
\set{(x,\, i)}[x = 0^n \,\lor\, x_i=1]
.}

Then
\m{
\Rx(g_1'),\, \bRx(g_1') \in \asO{\log n \dt \log\log n}
~~~\txt{and}~~~
\Dx(g_1') = n
.}
}

\prfstart

Solving \Sea\ is at least as hard as solving the \e{or} function and $\Dx(\Sea) = n$ follows from \clmref{c_Or}.

To solve \Sea\ in the model of \bRx\ (as well as in \Rx), we perform a binary search for the smallest non-zero coordinate of $x$ and output it if $x \ne 0^n$.
This requires solving $\log n$ instances of the $\lor$ function with error probability \asO{\dr1{\log n}}, which can be achieved via solving \asO{\log n \dt \log\log n} instances with error \asO1\ and the statement follows, again, from \clmref{c_Or}.

\prfend[\crlref{cr_Sea}]

Next we define the relational problem whose analysis will be the primary technical concern of this work -- the \e{approximate majority} relation, which is our example of $g_2$ such that $\Rx(g_2) \ll \bRx(g_2)$.

\ndefi[d_ApMa]{\ApMa, approximate majority problem}
{
Let $X=\OI^n$.
The following relational problem is called \e{approximate majority}:
\m{
\ApMa \deq~&
\set{x\in X}[|x| \le \fr n3]\times \set{0}\\
\cup~&
\set{x\in X}[\fr n3 < |x| < \fr{2n}3]\times \OI\\
\cup~&
\set{x\in X}[\fr{2n}3 \le |x|]\times \set{1}
.}
}

\sect[s_aff_GF2]{Some properties of affine subspaces in $\GF2[n]$}[Affine subspaces in GF₂ⁿ]

Parity-query protocols are naturally viewed as \e{partitions} of $\GF2[n]$ into \e{affine subspaces}.

Recall \defiref{d_RDx} and assume that $\Pi$ is a \Dx-protocol of complexity $k$.
It has a natural representation as a binary tree $T_\Pi$ of depth $k$:\ execution starts from the root, every non-leave vertex is marked with a parity query, the two edges leaving a node are marked by the complementary possible answers to the node's query and the leaves are marked with the answers that the protocol returns when reaching it.

The state of $\Pi$ at every moment -- that is, at each vertex of $T_\Pi$ (either a leaf or not) -- corresponds to an \e{affine subspace of $\GF2[n]$}:
Let $C\sbseq \OI^n$ be the set of input values corresponding to a vertex in $T_\Pi$ at depth $d\le k$, that is, for some 
\m{
(s_1,a_1)\dc (s_d,a_d) \in \pow{[n]}\times \OI
}
it holds that
\m{
C = \set{x\in \OI^n}[{\forall\: j\in [d]:\: \lxor_{i\in s_j}x_i=a_j}]
= \set{x\in \GF2[n]}[{\forall\: j\in [d]:\: \sum_{i\in s_j}x_i=a_j}]
.}
As a valid protocol state, $C$ must be non-empty.

Assume that the sequence $(s_1,a_1)\dc (s_d,a_d)$ is a shortest among those that define our $C$ in the above sense.
Then $s_1\dc s_d\in \GF2[n]$ must be linearly independent, as if, say, $s_1\dpl s_{t-1} = s_t$ for some $t\le d$, then $\lf[{\lxor_{i\in s_1}x_i}=a_1\dc \lxor_{i\in s_{t-1}}x_i=a_{t-1}\rt]$ implies $\lf[\lxor_{i\in s_t}x_i=a_t\rt]$ or its negation (or both), which contradicts at least one of the above assumptions.
Accordingly,
\f{
C' \deq \set{x\in \GF2[n]}[\sum_{i\in s_1}x_i=0\dc \sum_{i\in s_d}x_d=0]
}
is a subspace of $\GF2[n]$ of co-dimension $d$ and $C=C'+x_0$ for any $x_0\in C$, i.e., $C\lcure\GF2[n]$.

The leaves of the protocol tree $T_\Pi$ correspond to disjoint monochromatic (with respect to the computed function) affine subspaces of $\GF2[n]$ of co-dimension at most $k$, together forming a monochromatic partition of $\GF2[n]$ of size at most $2^k$ (attained if and only if $T_\Pi$ is complete).

In this work we will investigate some properties of affine subspaces in $\GF2[n]$.

\nlem[l_aff_coord]{Likely unfixed coordinates}
{
Let $n\ge 14$ and $C\lcure\GF2[n]$, $\Dim{C} \ge \dr{2n}3$.
Then there exists $J \sbs [n]$, $\sz{J} = \floor{\dr n3}$, such that for every $j\in J$, $X_j$ is unbiased when $X\sim \U[C]$ and
\m{
\forall\: D\lcure\GF2[n]:\:
\PR[j\unin J]{\txt{$x_j$ is fixed for all $x\in C\cap D$}}
\le 7 \tm \fr{\Dim{C} - \Dim{C\cap D}}n
.}
}

The claim is non-trivial owing to the universal quantifier in front of ``$D$'':\ that is, $J$ can depend only on $C$ and must be universal with respect to $D$.\,\fn
{
If $D$ were known, then $C\cap D$ would be known too, in which case letting $J$ contain as many as possible indices of non-fixed coordinates for $x\unin C\cap D$ and the rest of indices non-fixed for $x\unin C$ would satisfy the lemma promise. 
}
The claim is useful, as it will be applied later in the argument.
Finally, the claim is interesting, as it highlights a structural property that is special to affine spaces:\ if $C$ were allowed to be any large subset of $\GF2[n]$, then the analogous statement would be false (even if we drop the requirement for every $X_j$ to be unbiased under $X\sim \U[C]$).\,\fn
{
For sufficiently large $m\in \asT n$ with $\log m\in \NN$, pick $s_1\dc s_m\in \chs{[n-\log m]}{\floor{\dr n4}}$ uniformly at random and let
\m{
C \deq \bigcup_{i=1}^m
\set{x\in \GF2[n]}[x_{s_i} = \bar0,\, x_{\set{n-\log m+1\dc n}} = bin(i-1)]
,}
where $bin(i-1) \in \OI^{\log m}$ is the binary representation of $i-1$, then $\sz{C} \approx 2^{\dr{3n}4}$.
For large enough $n$ with high probability it will be the case that for every $J\in \chs{[n]}{\floor{\dr n3}}$ there is some $i_J\in [m]$, such that $\sz{J\cap s_{i_J}} \ge \dr{\sz{J}}5$.
Let $D_J \deq \set{x\in \GF2[n]}[x_{\set{n-\log m+1\dc n}} = bin(i_J-1)]$ -- clearly, this is an affine space of co-dimension $\log m$ and $\PR[j\unin J]{\txt{$x_j$ is fixed for all $x\in C\cap D$}} \ge \dr15$.
}

Let us have a closer look at the linear structure underlying the lemma.

Think about affine spaces in $\GF2[n]$ as being ``parametrised'' (and therefore represented) by the corresponding dual spaces.\fn
{
The dual linear space consists, precisely, of the ``known parities'' for the full set of elements of the corresponding primary affine space -- that is, in the context of a parity-query protocol $\Pi$, the dual space of (the affine subspace corresponding to) a vertex in $T_\Pi$ is the linear span of the queries marking the path from the root of the tree to that vertex.
}
On the one hand,
\m{
\wbr{C\cap D} = \spn{\wbr C \cup \wbr D} = \wbr C + \wbr D
,}
where ``$+$'' stands for element-wise addition (and the equality holds as we are working in $\GF2[n]$, where a linear combination is determined by the set of summands with non-zero coefficients).
On the other hand, $x_j$ is fixed for all $x\in C\cap D$ if and only if $e_j \in \wbr{C\cap D}$.
That is, the probability in the formulation of \lemref{l_aff_coord} can be rewritten as
\m[m_Pr_lem_coord]{
\PR[j\unin J]{e_j \in \wbr C + \wbr D}
}
and the lemma claims a rather strong upper bound on it (again, for any $D$).

Note that even though the set $J \sbs [n]$ in the statement can be made a function of the affine space $C$ (the corresponding statement would be logically equivalent to our lemma), constructing $J$ explicitly may be somewhat challenging:\ e.g., choosing any $J$ consisting of such $j$ that $e_{j} \nin \wbr C$ (that is, $X_j$ is unbiased in $X\unin C$) is unsuitable, as shown by letting $\wbr C$ be the linear span of $\set{e_1+e_2, e_1+e_3\dc e_1+e_{\dr n3+1}}$ and considering $J\deq \set{e_2, e_3\dc e_{\dr n3+1}}$ and any $D$ with $\wbr D$ containing $e_1$:\ in this case \bref{m_Pr_lem_coord} equals $1$ as $e_j \in \wbr C + \wbr D$ for every $j\in J$.\,\fn
{
This example can be easily generalised:\ e.g., allowing non-unique distances from $\wbr C$ to a very large set of weight-$1$ vectors that are nevertheless ``cancellable'' by some $D$ of small co-dimension.
Although in this work we do not need an explicit construction of $J$ in the statement of \lemref{l_aff_coord}, obtaining it might be interesting for its own sake.
}

\prfstart[\lemref{l_aff_coord}]

Towards contradiction, assume the opposite, that is, let $\forall J\in \chs{[n]}{\floor{\dr n3}}:\: \exists\: D_J\lcure\GF2[n]$, such that
\m{
\PR[j\unin J]{\txt{$x_j$ is fixed for all $x\in C\cap D_J$}}
> 7 \tm \fr{\Dim{C} - \Dim{C\cap D}}n
.}
Based on \bref{m_Pr_lem_coord} and the surrounding discussion, the above implies that for any such $J$ there exists $\wbr{D_J}\le \GF2[n]$ such that
\m{
\fr{\sz{\set{e_j}[j\in J] \cap \lf(\wbr C + \wbr{D_J}\rt)}}{\sz{J}}
> 7 \tm \fr{\Dim{C} - \Dim{C\cap D_J}}n
= 7 \tm \fr{\Dim{\wbr C + \wbr{D_J}} - \Dim{\wbr C}}n
,}
that is,
\m{
\sz{\set{e_j}[j\in J] \cap \lf(\wbr C + \wbr{D_J}\rt)}
> 2 \tm \lf(\Dim{\wbr C + \wbr{D_J}} - \Dim{\wbr C}\rt)
,}
as $\sz{J} = \floor{\dr n3}$ and $n\ge 14$.
As we can take such $\wbr{D_J}=\spn{w_1\dc w_{\Dim{\wbr{D_J}}}}$ that $\wbr C \cap \wbr{D_J} = \bar0$,
\m[m_lem_coord_contra]{
\forall J\in \chs{[n]}{\floor{\dr n3}}:\:
\exists\: w_1\dc w_k \in \GF2[n]:\:
\sz{\set{e_j}[j\in J] \cap \lf(\wbr C + \spn{w_1\dc w_k}\rt)}
> 2 \tm k
.}

Consider the following recursion, indexed by $i \ge 0$.
Let
\m{
C_0 \deq \wbr C
,}
then $\Dim{C_0} \le \dr n3$ by the lemma assumption.
As long as $\Dim{C_i} \le \dr{2n}3$, let $J_i\in \chs{[n]}{\floor{\dr n3}}$ be arbitrary, subject to
\m{
\set{e_j}[j\in J_i] \cap C_i = \es
:}
such $J_i$ necessarily exists, as otherwise $\Dim{C_i} \ge \Dim{\spn{\set{e_j}[j\in C_i]}} > \dr{2n}3$.
Apply \bref{m_lem_coord_contra} with $J_i$ taking place of $J$, then let
\m{
C_{i+1} \deq C_i + \spn{w_1\dc w_{k_i}}
,}
where $w_1\dc w_{k_i}$ are those guaranteed by \bref{m_lem_coord_contra}.
Then by the trivial induction,
\m[m_lem_coord_Dim1]{
\Dim{C_{i+1}} \le \Dim{C_i} + k_i \le \fr n3 + \sum_{j=0}^i k_j
.}

Let $\Gamma_i \deq \set{e_j}[j\in J_i] \cap C_{i+1}$ and $\Delta_i \deq \spn{w_1\dc w_{k_i}}$ (in the earlier notation this is $\wbr{D_{J_i}}$).
Then \bref{m_lem_coord_contra} guarantees that
\m{
\sz{\Gamma_i}
= \sz{\set{e_j}[j\in J_i] \cap \lf(C_i + \Delta_i\rt)}
\ge \sz{\set{e_j}[j\in J_i] \cap \lf(\wbr C + \Delta_i\rt)}
> 2 \tm k_i
,}
where the first inequality reflects the relation $\wbr C = C_0 \sbseq C_1 \ds[\sbseq] C_i$.
Trivially,
\m{
\Gamma_0\dc \Gamma_{i-1} \sbseq C_i
}
but
\m{
\Gamma_i \cap C_i = \es
,}
as $\Gamma_i \sbseq \set{e_j}[j\in J_i]$ while $\set{e_j}[j\in J_i] \cap C_i = \es$.
Accordingly, \pl[\Gamma_i] are pairwise disjoint and
\m[m_lem_coord_Dim2]{
\Dim{C_i}
\ge \Dim{\spn{\Gamma_0\ds[\cup] \Gamma_{i-1}}}
= \sum_{j=0}^{i-1} \sz{\Gamma_j}
> 2 \tm \sum_{j=0}^{i-1} k_j
,}
as \pl[\Gamma_i] consist of unit vectors from $\GF2[n]$.

Evidently, \bref{m_lem_coord_Dim1} and \bref{m_lem_coord_Dim2} disagree.
Indeed, let $i_0$ be the index of the last round in the above recursion, then it follows from the halting condition $[\Dim{C_{i_0}} > \dr{2n}3]$ and \bref{m_lem_coord_Dim1}:
\m{
\fr{2n}3
< \Dim{C_{i_0}}
\le \fr n3 + \sum_{j=0}^{i_0-1} k_j
,}
that is,
\m{
\Dim{C_{i_0}} < 2 \tm \sum_{j=0}^{i_0-1} k_j
,}
which is in contradiction with \bref{m_lem_coord_Dim2}, as required.

Finally, if $J \sbs [n]$ is such that for one or more indices $j\in J$ the coordinate $X_j$ is not unbiased when $X\sim \U[C]$ but the set $J$ satisfies the rest of lemma guarantees, then every such $j$ can be replaced by a coordinate that is ``free'' in $C$ without breaking the guarantees (note that a coordinate $X_i$ can only be either unbiased or fixed when $X\sim \U[C]$ for $C\lcure\GF2[n]$).

\prfend

We will often consider the behaviour of \e{deterministic} parity-query protocols (that is, of \Dx-protocols) with respect to random inputs coming from a known distribution of the following form.

\ndefi[d_muCt]{Distribution \muCt}
{
Assume that $C\lcure\GF2[n]$ and $0\le t\le \Dim{C} - \dr{2n}3$.
Denote by \muCt\ the following distribution of $X\in C$.
\itstart
\item
Choose $i_1$ uniformly at random from the set $J$ guaranteed by \lemref{l_aff_coord} with respect to $C$.
Let $C_1 \deq \set{x\in C}[x_{i_1} = 1]$.
\item
For each $2\le j\le t$ consecutively choose $i_j$ uniformly at random from the set $J$ guaranteed by \lemref{l_aff_coord} with respect to $C_{j-1}$ taking place of $C$ and let $C_j \deq \set{x\in C_{j-1}}[x_{i_j} = 1]$.
\item 
Output $X\sim \U[C_t]$.
\itend
}

In other words, $X\sim \muCt$ corresponds choosing $X\in C$ at uniform, subject to consecutive fixing $t$ random coordinates of the elements in $C$ to ``$1$'', each time selecting a ``likely unfixed'' coordinates according to \lemref{l_aff_coord}.

We will be arguing that a parity-query protocol $\Pi$ cannot distinguish well between the input distributions $\U[C]$ and \muCt, as long as $t$ is not too large with respect to the complexity of $\Pi$.
To that end we will use the following facts concerning a pair of affine spaces.

\lem[l_Crt]
{
Let $C,D\,\lcure\,\GF2[n]$, $\Dim{C} \ge \dr{3n}4$ and $t \le \fr n{14\Dim{\wbr D}}$.
Then
\m{
\muCt(D) \ge \fr 12\tm \U[C](D)
.}
}

The claim is non-trivial as the distribution $\muCt$ doesn't depend on $D$.

\prfstart[\lemref{l_Crt}]

Assume that $C\cap D\ne \es$ and $\Dim{\wbr D} \ge 1$ (otherwise the statement holds trivially), then $t \le \dr n{14}$.
Also assume $n\ge 14$ (otherwise $t=0$ and the statement holds trivially).
Note that for any $C'$ it holds that
\m{
\Dim{C'} - \Dim{C'\cap D}
= \Dim{\wbr C' + \wbr D} - \Dim{\wbr C'}
\le \Dim{\wbr D}
.}

Let us look at the procedure for generating $X\sim \muCt$, as given by \defiref{d_muCt}.
Denote $C_0 \deq C$, then the requirements of \lemref{l_aff_coord} are satisfied with respect to $C_0$ taking place of $C$ and it holds with respect to the choice of $i_1$ (from \defiref{d_muCt}) that
\m{
\PR[i_1]{\txt{$x_{i_1}$ is fixed for all $x\in C_0\cap D$}}
\le 7 \tm \fr{\Dim{\wbr D}}n
,}
where we view $i_1$ as a random variable.
For $2\le j\le t$, let $C_j \deq \set{x\in C_{j-1}}[x_{i_j} = 1]$, then the requirements of \lemref{l_aff_coord} are satisfied with respect to $C_j$, and so,
\m{
\PR[i_j]{\txt{$x_{i_j}$ is fixed for all $x\in C_{j-1}\cap D$}}
\le 7 \tm \fr{\Dim{\wbr D}}n
.}
By the union bound,
\m[m_lem_Crt_un]{
\PR[i_1\dc i_t]{\txt{any of $x_{i_j}$ is fixed for $x\in C_{j-1}\cap D$, $1\le j\le t$}}
\le 7t \tm \fr{\Dim{\wbr D}}n
\le \fr 12
,}
viewing $i_1\dc i_t$ as random variables that are distributed according to the procedure for generating $X\sim \muCt$ from \defiref{d_muCt}.

Now let us compare $\U[C](D)$ to $\muCt(D)$.
The former is the probability that $X\in D$ when $X\sim \U[C]$, which equals
\m{
2^{\Dim{C\cap D} - \Dim{C}}
.}
The latter is the probability that $X\in D$ when $X\sim \U[C_t]$ with respect to $C_t$ constructed in \defiref{d_muCt}, which equals
\m{
2^{\Dim{C_t\cap D} - \Dim{C_t}}
,\,\fnm
}
\fnt
{
Let $\Dim{\es} \deq -\infty$ (note that affine spaces do not need any ``special element'', like $\bar0$ for linear spaces, so empty affine spaces may be consistently allowed, thus making the whole concept closed under intersection).
}
where $[\Dim{C_t} = \Dim{C} - t]$ always (by the construction).
Accordingly, for a tuple $i_1'\dc i_t'\in [n]^t$ it can be the case that
\m{
\PR[X\sim \muCt]{X\in D}[i_1=i_1'\dc i_t=i_t']
< \U[C](D)
}
only if at least one of the (consecutive) choices ``$i_j=i_j'$'' fixes the corresponding coordinate in $x\in C_{j-1}\cap D$.
The probability of the latter is, according to \bref{m_lem_Crt_un}, at most \dr12, and therefore
\m{
\muCt(D) \ge \fr 12\tm \U[C](D)
,}
as required.

\prfend

\sect[s_ApMa]{Unambiguous parity-query complexity of approximate majority}[Unambiguity of appr.\ Maj]

\ntheo[t_ApMa]{Unambiguous parity-query complexity of \ApMa}
{
\m{
\bRx(\ApMa) \in \asOm{\sq n}
.}
}

Obviously, $\Rx(\ApMa) \in \asO1$, so the statement is interesting:
As we discussed earlier, the unambiguous regime in the model of parity queries defies some of the ``most intuitive'' lower-bound approaches, so the proof of \theoref{t_ApMa} will implement its own ad hoc strategy.

The analysis must distinguish between the cases of relational and functional problems, so we will investigate the behaviour of our $f$ in the region where $\ApMa$ would allow uncertainty:\ that is, on the input values of Hamming weight between $\dr{n}3$ and $\dr{2n}3$.
Intuitively, the transition from the ``$0$''-region to the ``$1$''-region of $\ApMa$ is hard to handle for an unambiguous protocol -- as opposed to an arbitrary randomised one.
An unambiguous protocol has to ``adhere to'' its own choices of the answers in the uncertainty region, while usual randomised protocols may answer inconsistently there.

As \Rx-protocols of complexity $k$ are convex combinations of \Dx-protocols, which, in turn, partition the input space $\GF2[n]$ into at most $2^k$ affine subspaces whose elements receive the same answer, our proof will be based technically on bounding the ability of large affine subspaces to discriminate input values, based on their Hamming weight.

\prfstart[\theoref{t_ApMa}]

Let $k(n) \in \asO{\bRx(\ApMa)}$ be the parity-query complexity of unambiguously solving \ApMa\ with error at most $\dr 1{20}$ and denote by $f:\OI^n \to \OI$ some \ApMa-consistent function of that complexity -- that is, $f$ takes value ``$0$'' on the inputs of Hamming weight at most $\dr{n}3$ and value ``$1$'' on the inputs of Hamming weight at least $\dr{2n}3$.
Assume without loss of generality that $k(n)$ is monotone non-decreasing and
\m{
\PR[{\U[\OI^n]}]{f(X) = 0} \ge \fr 12
}
(otherwise replace $f(x)$ by $1-f(\lnot x)$, where ``$\lnot$'' stands for the bit-wise negation).

Let $n_0 \deq n$, $C_0 \deq \GF2[n_0]$, $f_0 \deq f$ and consider the following recursion, indexed by $j \ge 0$.

We will make sure that for every $j$ throughout the recursion:\ $n_j \le n_{j-1}$, $C_j \lcure \GF2[n_j]$, $f_j:C_j \to \OI$ is a sub-function of $f$ (namely, a restriction of $f$ to certain affine subspace of dimension $n_j$) and $\PR[{\U[C_j]}]{f_j(X) = 0} \ge \dr 12$.
Let $t_j \deq \floor{\fr{n_j}{14\tm k(n_j)}}$, $\mu_j$ be the distribution $\mu_{C_j}^{(t_j)}$ and $\nu_j\deq \fr{\U[C_j]+\mu_j}2$.

As long as $\Dim{C_j} \ge \dr{3n_j}4$ and $\sum_{\ell<j} t_\ell < \dr {2n}3$, let $\Pi_j$ be a parity-query deterministic protocol of complexity at most $k(n_j)$ that computes $f_j$ with error at most $\dr 1{20}$ with respect to the distribution $\nu_j$.
Then it has error at most $\dr 1{10}$ with respect to both $\U[C_j]$ and $\mu_j$.
As $\PR[{\U[C_j]}]{f_j(X) = 0} \ge \dr 12$, the protocol answers ``$0$'' with probability at least $\dr 12 - \dr 1{10} = \dr 25$ with respect to $X\sim \U[C_j]$.

Denote by $\Cl D_j$ the family of affine subspaces of $\GF2[n_j]$ corresponding to the ``$0$''-marked leaves of $\Pi_j$ ($D\in \Cl D_j$ are pairwise disjoint, each of co-dimension at most $k(n_j)$).
Then by \lemref{l_Crt}:
\m{
\sum_{D\in \Cl D_j} \mu_j(D) \ge \fr 12\tm \sum_{D\in \Cl D_j} \U[C_j](D) \ge \fr 15
.}
Accordingly,
\m{
\PR[X\sim \mu_j]{f_j(X) = 1}[X\in \cup_{D\in \Cl D_j} D]
= \fr{\PR[\mu_j]{\txt{$\Pi_j$ errs and $X\in \cup_{D\in \Cl D_j} D$}}}
{\sum_{D\in \Cl D_j} \mu_j(D)}
\le 5\tm \PR[\mu_j]{\txt{$\Pi_j$ errs}}
\le \fr 12
}
and there must exist some $D_j \in \Cl D_j$, such that $C_j\cap D_j\ne \es$ and
\m[m_t_ApMa_mui]{
\PR[X\sim \mu_j]{f_j(X) = 1}[X\in D_j] \le \fr 12
.}

Here again (like in the proof of \lemref{l_Crt}), let us view $i_1\dc i_{t_j}$ as random variables that accompany the generation of $X\sim \mu_j = \mu_{C_j}^{(t_j)}$ according to \defiref{d_muCt}.
Then
\m{
&\PR[X\sim \mu_j]{f_j(X) = 1}[X\in D_j]\\
&\qd = \sum_{i_1'\dc i_{t_j}'}
\PR{i_1=i_1'\dc i_{t_j}=i_{t_j}'}
\tm \PR[{X\sim \U[C_j]}]{f_j(X) = 1}[{X\in D_j,\, X_{i_1'}\ds[=] X_{i_{t_j}'} = 1}]
,}
and therefore for some values $i_1'\dc i_{t_j}'\in [n_j]$ it holds that $\PR{i_1=i_1'\dc i_{t_j}=i_{t_j}'} > 0$ and
\m[m_t_ApMa_ips]{
\PR[{X\sim \U[C_j]}]{f_j(X) = 1}[{X\in D_j,\, X_{i_1'}\ds[=] X_{i_{t_j}'} = 1}]
\le \PR[X\sim \mu_j]{f_j(X) = 1}[X\in D_j]
\le \fr 12
,}
where the last inequality is \bref{m_t_ApMa_mui}.
These $i_1'\dc i_{t_j}'$ are distinct and every $X_{i_\ell'}$ is unbiased under $X\sim \U[C_j]$ by the construction of $\mu_j$ (\defiref{d_muCt}) and the guarantees of \lemref{l_aff_coord}.

Let
\m[P]{
&L_j \deq \set{i_1'\dc i_{t_j}'};\\
&\wbr{L_j} \deq [n_j]\smin L_j;\\
&\wht{C_j} \deq \set{x\in C_j \cap D_j}[{x_{i_1'}\ds[=] x_{i_{t_j}'} = 1}];\\
&n_{j+1} \deq n_j-t_j;\\
&C_{j+1} \deq \set{x_{\wbr{L_j}}}[x\in \wht{C_j}]
~~(\txt{the element-wise projection of $\wht{C_j}$ on $\wbr{L_j}$});\\
&\forall\: x\in \wht{C_j}:\: f_{j+1}(x_{\wbr{L_j}}) \deq f_j(x)
~~(\txt{this defines $f_{j+1}:C_{j+1} \to \OI$})
.}
As $C_j \lcure \GF2[n_j]$ by the induction hypothesis, $\wht{C_j} \lcure \GF2[n_j]$ too and $C_{j+1} \lcure \GF2[n_{j+1}]$.
Obviously, $f_{j+1}$ is a sub-function of $f_j$ (and therefore of $f$) and
\m[m_t_ApMa_f_12]{
\PR[{X\sim \U[C_{j+1}]}]{f_{j+1}(X) = 0}
= 1 - \PR[{X\sim \U[C_j]}]{f_j(X) = 1}[{X\in D_j,\, X_{i_1'}\ds[=] X_{i_{t_j}'} = 1}]
\ge \fr 12
,}
according to \bref{m_t_ApMa_ips}.
So, the conditions that we assumed to hold in the beginning of the \ord[j] iteration will also hold for $j+1$, and thus our recursion can sustain itself.

Let $j_0$ be the index of the last protocol $\Pi_j$ considered in the recursion, then it follows from the halting condition that either
\m[m_t_ApMa_case1]{
\Dim{C_{j_0+1}} < \fr{3n_{j_0+1}}4
}
or
\m[m_t_ApMa_case2]{
\sum_{\ell=0}^{j_0} t_\ell \ge \fr {2n}3
.}

By \bref{m_t_ApMa_f_12} with respect to round $j_0$ of the recursion, $\PR[{\U[C_{j_0+1}]}]{f_{j_0+1}(X) = 0} \ge \dr 12$.
On the other hand, the affine subspace $C_{j_0+1}$ is the outcome of a series of restrictions imposed upon $C_0 = \GF2[n_0]$, which, in particular, have constrained to ``$1$'' the values of $\sum_{\ell=0}^{j_0} t_\ell$ previously unfixed coordinates.
Then \bref{m_t_ApMa_case2} would imply that $f_{j_0+1}:C_{j_0+1} \to \OI$ is a restriction of $f$ to input values of Hamming weight at least $\dr {2n}3$, and $f_{j_0+1} \equiv 1$ would follow, as $f$ is assumed to be a \ApMa-consistent function.
Accordingly, \bref{m_t_ApMa_case1} holds and \bref{m_t_ApMa_case2} doesn't, that is, $\sum_{\ell\le j_0} t_\ell < \dr {2n}3$ and $\Dim{C_{j_0+1}} < \fr{3n_{j_0+1}}4$.

By construction, $\forall\: j \le j_0$ it holds that $n_{j+1} = n - \sum_{\ell=0}^{j} t_\ell$ and
\m{
\Dim{C_{j+1}}
& \ge \Dim{C_j} - \Dim{\wbr{D_j}} - t_j\\
& \ge \Dim{C_j} - k(n_j) - t_j\\
& \ge n - \sum_{\ell=0}^j k(n_\ell) - \sum_{\ell=0}^j t_\ell
= n_{j+1} - \sum_{\ell=0}^j k(n_\ell)
,}
and so,
\m{
\sum_{\ell=0}^{j_0} k(n_\ell)
\ge n_{j_0+1} - \Dim{C_{j_0+1}}
> \fr{n_{j_0+1}}{4}
.}
On the other hand,
\m{
n_{j_0+1} = n - \sum_{\ell=0}^{j_0} t_\ell > \fr n3
,}
therefore,
\m{
\sum_{\ell=0}^{j_0} k(n_\ell)
> \fr n{12}
> \fr18 \tm \sum_{\ell=0}^{j_0} t_\ell
}
and there exists $\ell_0 \le j_0$ such that
\m{
k(n_{\ell_0})
> \fr18 \tm t_{\ell_0}
= \fr18 \tm \floor{\fr{n_{\ell_0}}{14\tm k(n_{\ell_0})}}
.}
Therefore, $k(n_{\ell_0}) \in \asOm{\sq{n_{\ell_0}}} = \asOm{\sq n}$ and the result follows.

\prfend

That is, imposing the restriction of unambiguity can turn a problem that is easy for parity queries with randomness into a hard problem.

\crl[cr_ApMa]
{
\m{
\Rx(\ApMa) \in \asO1
~~~\txt{and}~~~
\bRx(\ApMa),\, \Dx(\ApMa) \in \asOm{\sq n}
.}
}

\prfstart

An \Rx-protocol that queries constant number of bits at random locations can answer $\ApMa$ with arbitrarily small constant-bounded error probability.

\prfend[\crlref{cr_ApMa}]

It is, in fact, not hard to see that $\Dx(\ApMa) \in \asOm n$.

\sect[s_Concl]{Conclusions}

We believe that further investigation of the concept of unambiguity is likely to offer a peerless insight into the phenomenon of computational randomness in itself.
As discussed in \sref{s_intro}, restricting one's curiosity to the deterministic and the randomised regimes alone may be insufficient for an adequate understanding of computational randomness.

For the sake of speculation, next we pose two questions related to computational randomness and then use them as a basis for a hypothetical ``road map'' from the results of this work towards better understanding of randomness.

Efficient deterministic parity-query protocols are partitions of the input space $\GF2[n]$ into a (relatively) small number of same-answer \e{affine subspaces}; accordingly, a function with an efficient deterministic parity-query protocol must be constant on some large affine subspaces of $\GF2[n]$.
Randomised protocols are convex combinations of deterministic ones, so there seems to be no need for a function with an efficient randomised parity-query protocol to be constant on a large affine subspace of $\GF2[n]$.
Nevertheless, in all known cases a function over $\GF2[n]$ with an efficient randomised parity-query protocol \e{is constant over some large affine subspace}.

\nquest[concl_q_1]{\cite{KLGY21_Mod}}{Is every $f:\: \GF2[n] \to \OI$ with an efficient randomised parity-query protocol constant on a large affine subspace of $\GF2[n]$?}

Similarly, while every deterministic bipartite protocol with small communication cost corresponds to a partition of the input space $\OI^n\times \OI^n$ into a small number of same-answer \e{combinatorial rectangles}, there seems to be no immediate reason for a function with an efficient randomised protocol to be constant on a large sub-rectangle of $\OI^n\times \OI^n$.
Nevertheless, all known examples of such functions do have a large rectangle over which they are constant.

\nquest[concl_q_2]{\cite{CLV19_Equa}}{Is every $f:\: \OI^n\times \OI^n \to \OI$ with an efficient randomised communication protocol constant on a large combinatorial rectangle?}

If the answer to \questref{concl_q_1} were affirmative, this would imply some variation of \theoref{t_ApMa}, as it is easy to see that every large affine subspace of $\GF2[n]$ contains both elements of Hamming weight less that $\dr n3$ and those of Hamming weight more than $\dr {2n}3$.
Accordingly, this work is a step towards answering \questref{concl_q_1}.

Any \Rx-protocol for a function $f:\: \GF2[n] \to \OI$ can be emulated by a bipartite randomised communication protocol of roughly the same complexity for the function $F(x,y) \deq f(x\lxor y)$ -- obviously, only requires a very special and rather restricted (but well-defined) type of randomised bipartite communication protocols.
Any function $F(\dt,\dt)$ that can be solved efficiently by one of such restricted protocols necessarily corresponds, in the above sense, to some $f:\: \GF2[n] \to \OI$ with an efficient randomised parity-query protocol; the affirmative answer to \questref{concl_q_1} would imply that $f$ is constant on a large affine subspace $C \lcure \GF2[n]$.
A large same-answer affine subspace for $f$ corresponds to a large same-answer combinatorial rectangle for $F$ itself:
\m{
C = C' + c_0 = C' + \lf( C' + c_0 \rt)
}
for any $c_0\in C$, thus the rectangle $r_C \deq C' \times ( C' + c_0 )$ satisfies $\set{x \lxor y}[(x,y)\in r_C] = C$ and is monochromatic with respect to $F$.
That is, the affirmative answer to \questref{concl_q_1} could be reinterpreted as affirming a special case of \questref{concl_q_2}.

A well-known structural question is this:
\vsp[-1]\cent{
\e{Is \BPP\ inside $\Pp^{\NP}$ in communication complexity?}
}\vsp[-1]
Here ``\BPP'' stands for the family of all bipartite functions (with product domains) that are efficiently computable by randomised communication protocols and ``$\Pp^{\NP}$'' is the natural communication-complexity analogue of the corresponding class in computational complexity.
It is known~\cite{IW10_Com} that every function with an efficient $\Pp^{\NP}$-protocol is constant on a large combinatorial rectangle, so ``$\BPP \sbseq \Pp^{\NP}$'' would imply the affirmative answer to \questref{concl_q_2}.

And so on\ldots

\toct{Acknowledgements}

\sect*{Acknowledgements}

I am indebted to Shalev Ben-David, who first presented to me the thrilling idea of unambiguous complexity, and to Alex Samorodnitsky, who helped me to gain some familiarity with the fanciful realm of binary linear spaces.

\toct{References}

\end{document}